\documentclass[aps,rmp,reprint,twocolumn,unsortedaddress,showpacs,longbibliography]{revtex4}

\usepackage{graphicx}
\usepackage{hyperref}
\usepackage{color}
\definecolor{mybo}{cmyk}{0,0,0.10,0}

\newcommand{\tabincell}[2]{\begin{tabular}{@{}#1@{}}#2\end{tabular}}

\begin{document}

\title{Hybrid quantum circuits: Superconducting circuits interacting with other quantum systems}

\author{Ze-Liang Xiang}
\email{zlxiang@riken.jp}
\affiliation{Department of Physics and State Key Laboratory of Surface Physics, Fudan University, Shanghai 200433, China}
\affiliation{Advanced Science Institute, RIKEN, Saitama 351-0198, Japan,}

\author{Sahel Ashhab}
\email{ashhab@riken.jp}
\affiliation{Advanced Science Institute, RIKEN, Saitama 351-0198, Japan,}
\affiliation{Department of Physics, The University of Michigan, Ann Arbor, Michigan 48109, USA}

\author{J. Q. You}
\email{jqyou@fudan.edu.cn}
\affiliation{Department of Physics and State Key Laboratory of Surface Physics, Fudan University, Shanghai 200433, China,}
\affiliation{Beijing Computational Science Research Center, Beijing 100084, China,}
\affiliation{Advanced Science Institute, RIKEN, Saitama 351-0198, Japan}

\author{Franco Nori}
\email{fnori@riken.jp}
\affiliation{Advanced Science Institute, RIKEN, Saitama 351-0198, Japan,}
\affiliation{Department of Physics, The University of Michigan, Ann Arbor, Michigan 48109, USA}

\begin{abstract}
Hybrid quantum circuits combine two or more physical systems, with the goal of harnessing the advantages and strengths of the different systems in order to better explore new phenomena and potentially bring about novel quantum technologies. This article presents a brief overview of the progress achieved so far in the field of hybrid circuits involving atoms, spins and solid-state devices (including superconducting and nanomechanical systems). How these circuits combine elements from atomic physics, quantum optics, condensed matter physics, and nanoscience is discussed, and different possible approaches for integrating various systems into a single circuit are presented. In particular, hybrid quantum circuits can be fabricated on a chip, facilitating their future scalability, which is crucial for building future quantum technologies, including quantum detectors, simulators, and computers.
\end{abstract}


\date{\today}
\maketitle
\tableofcontents

\section{Introduction}\label{sec:intro}

The field of quantum information processing is attracting considerable interest, and scientists in a variety of disciplines have devoted intense effort to the realization of quantum information principles, technologies, and algorithms~\cite{Bennett:2000,DiVincenzo:2000,Nielsen:2000,Schleich:2008,Stolze:2008,Georgescu:2012}. The most advanced experimental demonstrations of controllable quantum coherent systems include trapped ions and atoms~\cite{Buluta:2011,Bloch:2008,Blatt:2008}, spins~\cite{Buluta:2011,Hanson:2007,Hanson:2008a}, and superconducting circuits~\cite{Makhlin:2001,You:2005,Wendin:2007,Clarke:2008,You:2011,Zagoskin:2011,Buluta:2011}.

The properties of atoms have been studied extensively over the past century. Atoms have stable energy levels that can be used to represent the different states of qubits. In addition, the coherence times of isolated atoms are long because of their weak interaction with the surrounding environment~\cite{Lukin:2003,Blatt:2008}.

Spins are another promising candidate for future quantum technologies~\cite{Hanson:2008a}. Impurity spins can behave as qubits and can be used to store or process quantum information. For example, phosphorous impurities in silicon~\cite{Kane:1998,Morello:2010} and nitrogen-vacancy (NV) color centers in diamond~\cite{Wrachtrup:2006,Doherty:2012} possess good coherence properties, which allow long storage times. Furthermore, rapid progress has been made with quantum dots~\cite{Loss:1998,Hanson:2007,Zwanenburg:2012}, which can be fabricated on a chip and controlled relatively easily using electric signals.

Remarkable progress has also been made on other systems, such as superconducting (SC) circuits. These SC circuits promise good scalability~\cite{Ashhab:2008,Helmer:2009,Galiautdinov:2012,Harris:2012} and allow robust control, storage and readout due to their strong interaction with external fields.

Table~\ref{Table1} compares different systems used as qubits. Each system has its own advantages and disadvantages [see, e.g., \textcite{Buluta:2011}, for a recent overview]. Macroscopic systems, such as SC qubits, offer flexibility, tunability, scalability, and strong coupling to external fields, but have relatively short coherence times ($\lesssim0.1$ ms) and in general are not identically reproducible. On the other hand, microscopic systems, such as atoms and spins, are given by nature and can easily be made as identical qubits with long coherence times ($\gtrsim1$ ms), but they operate slowly because of their weak coupling to external fields, and have limited scalability (i.e., it is difficult to individually control many atoms working as qubits). A promising idea pursued by various groups at the moment is to combine these different systems and build new hybrid quantum structures that would inherit only the advantages of each one of the different components~\cite{DiVincenzo:2011,Wallquist:2009,Duty:2010}.

The most actively pursued type of hybrid quantum circuits (HQCs) uses SC qubits because of their strong coupling to external fields and ease of control. Indeed, rapid progress has been achieved with SC circuits, and many groups are devoting efforts to find a good model to build a hybrid quantum circuit. The long-time storage of the quantum information can be implemented in microscopic systems, such as atoms or spins~\cite{Blencowe:2010,Duty:2010}. This situation is analogous to that in classical computers, which combine electronic circuits and magnetic drives to achieve fast processing and robust long-time information storage, respectively.

HQCs involving SC circuits and nanomechanical resonators (NAMRs) are another emerging type of hybrid quantum systems~\cite{Armour:2002}. Recently, many studies have been devoted to this new and exciting subject for its promising applications in future quantum devices~\cite{Armour:2002, Blencowe:2004,Cleland:2004,Lahaye:2009} and the possibility of observing the quantum-to-classical transition in such a macroscopic object~\cite{Lahaye:2004,Wei:2006}. With current technology, NAMRs can be fabricated with SC circuits on a small chip, and have high quality factors and high vibrational mode frequencies~\cite{O'Connell:2010}. Through an effective coupling between the NAMR and the SC circuit, the NAMR can be considered as a``cavity'' to study cavity quantum electrodynamics (QED) and also can be cooled down to explore the transition from classical to quantum mechanics.

\begin{table*}[tbp]
\begin{center}\label{Table1}
\colorbox{mybo}{
\begin{tabular}{@{}p{3.0cm}p{0.1cm}|p{0.1cm}p{3.9cm}p{0.1cm}p{3.75cm}p{0.1cm}p{2.2cm}p{3.2cm}@{}}
\multicolumn{8}{l}{\rule[-3mm]{0mm}{8mm}{\bfseries TABLE I. ~Comparison between different systems used as qubits.}}\\
\hline\hline
\rule[-3mm]{0mm}{8mm} & & & {\it Atom, molecule, ion} && {\it Electron spin} && {\it Nuclear spin} & {\it Superconducting qubit} \\
\hline
{\bf Size} & & & $\sim 10^{-10}$ m && $\sim10 ^{-10}$ m (impurities) $~~$ $\sim10^{-8}$ m (quantum dot)\tablenotemark[1] && $< 10^{-10}$ m\tablenotemark[1] & $\sim 10^{-6}$ m\\
{\bf Energy gap} & & & $10^5$--$10^6$ GHz,  $~~~~~~~~~~~~~~~~~~~~$ $\sim$ GHz (Rydberg atoms) && 1--10 GHz && 1--10 MHz & 1--20 GHz\\
{\bf Frequency range} & & & Optical, microwave && Microwave && Microwave & Microwave\\
{\bf Operating$~~~~~~~$ temperature} & & & nK to $\mu$K && $\sim100$ mK (quantum dot)$~~$ room temp. (NV center) && $\sim$ mK & $\sim$ 10 mK\\
{\bf Single-qubit gate$~~$ operation time $\tau_1$} & & & $\sim~\mu$s (atom)$~~\quad\quad\quad\quad\quad\quad$ $\sim$ 50 ps (ion) & & $\sim$ 10 ns && $>10~\mu$s & $\sim$ 1 ns\\
{\bf Two-qubit gate$~~~$ operation time $\tau_2$} & & & $\sim~\mu$s (atom)$~~\quad\quad\quad\quad\quad\quad$ $\sim~100~\mu$s (ion) & & $\sim$ 0.2 ns && $\sim$ 10 ms & $\sim$ 10--50 ns \\
{\bf Coherence time $T_2$} & & & ms to s && ms to s && $\sim$ s & $\sim$ 10--100 $\mu$s \\
{\bf $T_2/\tau_1$} & & & $10$--$10^4$ && $10^5$--$10^8$ && $10^6$ & $10^4$--$10^5$ \\
{\bf Coupling type} & & & Electric or magnetic && Magnetic or electric && Magnetic & Electric or magnetic \\
{\bf Coupling strength with the cavity} & & & $<$ kHz ($B$-field), $~~~~~~~~~~~~~~~~~~$ $\sim$ 10 kHz ($E$-field),  $~~~~~~~~~~~~~~$ $\sim$ 10 MHz (Rydberg atoms) && $\sim$ 100 Hz (impurities) $~~~~$ $>$ MHz (quantum dot) && $\sim0.1$ Hz & $\sim$ 0.1--1 GHz \\
\hline\hline
\end{tabular}}
\end{center}
\tablenotetext[1]{$\,$Regarding the size of the electron and nuclear spins, as these are carried by pointlike particles, the size might be $\lesssim10^{-15}$ m. However, the more relevant quantity in this context is the size contributing to the interaction. In particular, the interaction is typically determined by the overlap integral of the coupled objects, and therefore the relevant size is the spatial extension of the wavefunction.}
\end{table*}

By far the most successful hybrid system involving SC qubits is the combination of a cavity and SC qubits. This hybrid system, often called circuit QED, has been reviewed and widely used in quantum technologies in the past~\cite{You:2005,Schoelkopf:2008}. Here we focus on more complex hybrid systems involving ``SC qubits plus a cavity'' (a hybrid system itself) {\it interacting with other quantum systems}, including atoms (both natural and artificial) as well as ions and various other quantum systems. Namely, the hybrid circuits we will be focusing on here are composed of various types of quantum systems interacting with a (hybrid) circuit QED system.

In this review, we first overview different basic elements needed to build HQCs, and then we highlight the progress achieved so far integrating these systems.

\section{Elements in hybrid quantum circuits}\label{sec:diff}

\subsection{Atoms}\label{ssec:atom}

Atomic systems (including neutral atoms, polar molecules, and ions) have been studied for a long time, and are promising systems for future quantum technologies. 

\begin{figure*}[tbp]
\includegraphics[width=5.75in]{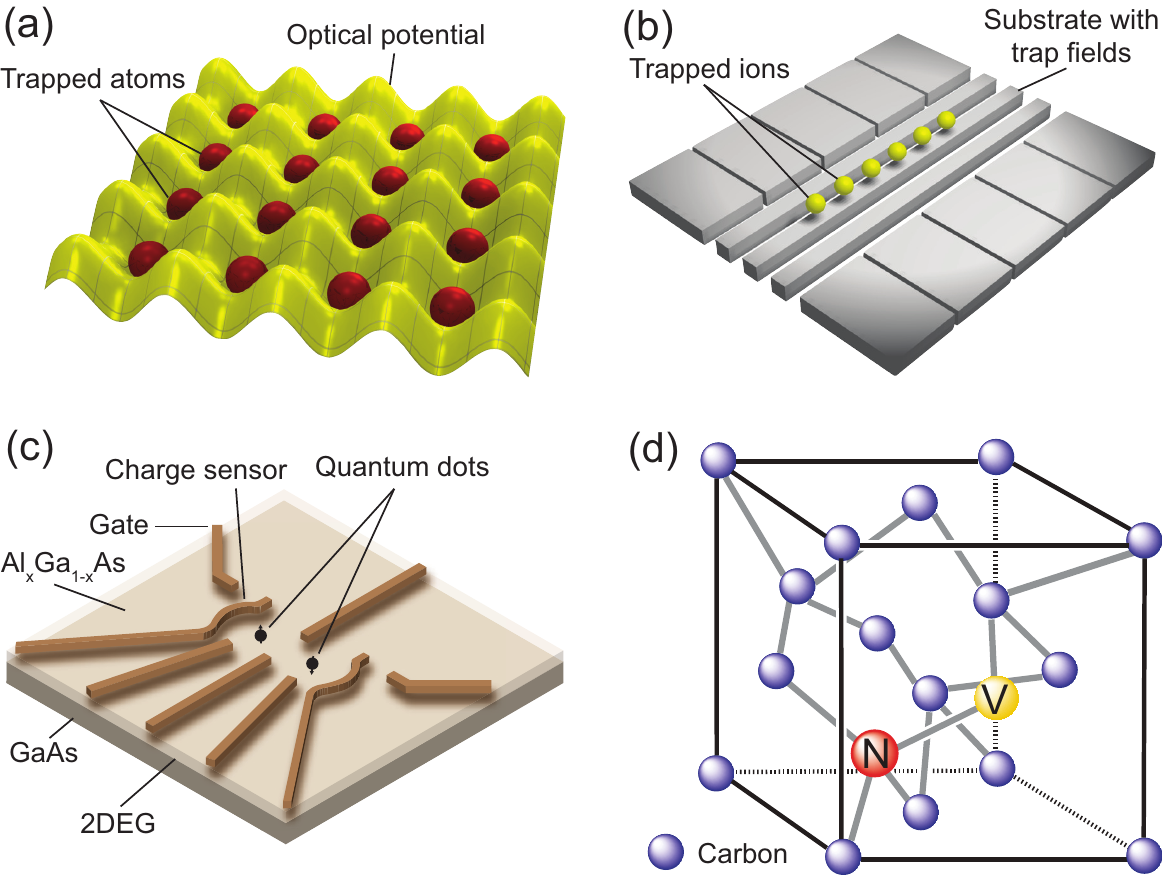}
\caption{(color online). Different types of qubits: (a) atoms trapped in an optical lattice; (b) a planar trap of ions; (c) an electrostatically-defined quantum dot; and (d) a nitrogen-vacancy (NV) center in diamond. (a) and (b) are adapted from \textcite{Buluta:2011}.}
\label{ATSP}
\end{figure*}

Because of the weak interaction with the environment, atomic systems can achieve long coherence times, which is useful for storing quantum information. In general, optical pumping and cooling, electromagnetic radiation, and laser-induced fluorescence are utilized to initialize, manipulate, and measure the qubit encoded in the atomic levels~\cite{Lukin:2003}. Recently, much progress was achieved studying atoms trapped in optical lattices; see Fig.~\ref{ATSP}(a). In this ultra-low-temperature system, atoms are trapped in microscopic arrays created by laser beams, and can be precisely manipulated for eventual use in future quantum devices [see, e.g., \textcite{Wieman:1999} and \textcite{Bloch:2008}] and simulators [see, e.g., \textcite{Jaksch:2005}, \textcite{Buluta:2009}, \textcite{Lewenstein:2007}, and \textcite{Georgescu:2011}]. 

However, because atoms interact with each other weakly, achieving many-qubit entangled states or quantum gate operations becomes a major challenge. As a way to overcome this problem, Rydberg states of atoms~\cite{Saffman:2010} [or polar molecules~\cite{Micheli:2006}] with their large dipole-dipole interactions have been explored for the realization of quantum gate operations. 

Rydberg states occur when an atom is excited such that one of the electrons moves into a high principal quantum number orbital. Such Rydberg atoms are much more sensitive to external fields and possess very large electric dipole moments. Similar to the Coulomb blockade and the photon blockade, the Rydberg blockade, which prevents atoms from being excited to a Rydberg state if there is another Rydberg atom nearby, can be used to entangle two atoms located in two separate optical dipole traps and implement effective two-qubit gate operations~\cite{Saffman:2010}. 

Ions are another prospective type of qubits with long coherence times~\cite{Cirac:2000,Blatt:2008}. Because they are charged, interactions among ions via Coulomb repulsion are much stronger than those among atoms. This property facilitates the realization of two- or multiqubit operations. Ions can also be cooled by laser beams and trapped by electrical or magnetic fields, and they can be manipulated with high precision, see Fig.~\ref{ATSP}(b). Generally, hyperfine or Zeeman sublevels, the ground and excited states of an optical transition, and the collective motion of ions can be used to encode quantum information with long lifetimes $>20$~s, $>1$~s, and $<100$~ms, respectively. The single-qubit gate time of trapped ions is around 50 ps, while the two-qubit gate time is around hundreds of microseconds~\cite{Buluta:2011}. The manipulation of ions can be achieved using the same methods used for atoms.

In atomic systems, the coherence times are very long. However, compared to other systems, initialization, operation and measurement times are also very long. Thus, combining an atomic system (with long coherence times) with another system that allows fast operations has emerged as a possible way to construct novel devices benefiting from the advantages of two seemingly different systems.

\subsection{Spins}\label{ssec:spin}

Spins can also serve as qubits and store or process quantum information~\cite{Hanson:2008a}. In general, two kinds of spins are used in quantum computation: electron spin and nuclear spin. Both types can interact with the electric or magnetic fields of a photon.

One can trap atomic gases and utilize electron spins to store quantum information. However, the techniques of trapping and cooling are rather complicated. Alternatively, by integrating dopants into a solid-state host material, large arrays of spin qubits can more easily be assembled in experiment.

Spins in solids generally fall into two classes: quantum dots and atomic impurities. Quantum dots are small nanostructures where electrons are trapped in a potential well and have discrete energy levels~\cite{Loss:1998,Hanson:2007,Zwanenburg:2012}; see Fig.~\ref{ATSP}(c). These come in several forms. One is electrostatically defined quantum dots, where the distribution of electrons is controlled by voltages on lithographically fabricated metallic gates. Another form is self-assembled quantum dots, where electrons are confined by the artificial potential produced by a semiconductor growth process. In both types of quantum dots, by employing electrical or optical control, the manipulation, storage and readout of the qubit have been demonstrated in experiment, with typical gate times $\sim 200$ ps~\cite{Hanson:2008a,Petta:2005}. As reported by \textcite{Nowack:2007}, the coupling strength between the electric field and a single electron spin in a quantum dot can reach $\sim 5$ MHz, which is much larger than the coupling strength ($\sim 100$ Hz) between the external magnetic field and the electron spin in an impurity~\cite{Schoelkopf:2008}. However, the interaction between the spin and its surrounding spin bath (mainly nuclear spins) has so far limited coherence times to hundreds of microseconds or less~\cite{Bluhm:2011}.

Atomic impurities, such as phosphorus in silicon~\cite{Kane:1998,Morello:2010,Pla:2012}, Er$^{3+}$ ions in ${\rm Y_2SiO_5}$ crystal~\cite{Guillot-Noel:2006}, and NV color centers in diamond~\cite{Doherty:2012,Wrachtrup:2006}, can be conveniently integrated in solid-state devices and can have nuclear spins, electron spins, or both. Phosphorus impurities in silicon involve both electron and nuclear spins, and they possess excellent coherence properties [both have spin coherence times $T_2>1$ s~\cite{Morton:2008,Simmons:2011,Tyryshkin:2012,Steger:2012}]. Gate times are on the order of 1 ns. Using microwave and radio-frequency pulses, spins can implement quantum information processing, especially storage, in a conventional semiconductor material. Also, these spins can be controlled by strain~\cite{Weiler:2011} or via the Stark shift~\cite{Bradbury:2006}. Electron spins of Er$^{3+}$ ions doped in a crystal also have long coherence times ($>1$ ms). Either optical or microwave photon states can be mapped into spin states of Er$^{3+}$ ions by different energy-level transitions~\cite{Guillot-Noel:2006,Bushev:2011}. Thus, Er$^{3+}$ ions are a good candidate for use as a quantum interface between microwave and optical photons.

An NV center in diamond consists of a substitutional nitrogen (N) atom replacing a carbon atom and neighboring one vacancy (V). In such centers, both electron and nuclear spins can be used in quantum technologies and also exhibit long coherence times ($\sim1$ ms for the electron spin and $>1$ s for the nuclear spin; the single-qubit gate time is $\sim10$ ns for electron spins and $>10~\mu$s for nuclear spins) in a wide temperature range, even at room temperature~\cite{Gaebel:2006,Hanson:2008b,Balasubramanian:2009,Neumann:2010,Maurer:2012,Sar:2012}. Importantly, these NV centers can be used to detect weak magnetic fields~\cite{Maze:2008,Taylor:2008} and electric fields~\cite{Dolde:2011} at room temperature, instead of using low-temperature sensors. 

The electronic ground state of the NV center is a spin-1 triplet, $|m_s\rangle$ with $m_s=0,\pm1$. In the absence of a magnetic field, the $m_s=\pm 1$ sublevel is two-fold degenerate and the resonance transition frequency between the $m_s=0$ and $m_s=\pm 1$ sublevels is 2.87 GHz, which nicely matches the microwave frequency regime of SC qubits. This zero-field splitting in the NV center is caused by the reduction of the spin's rotation symmetry in the crystal. In the presence of a magnetic field, the $m_s=\pm 1$ sublevel is split into two levels, which makes it possible to address these two levels individually. NV centers also possess transitions from the electronic ground state to an excited electronic state, and the corresponding transition frequency is in the optical regime. These transitions can be used for the initialization and readout of the quantum state. Therefore, by using both laser and microwave fields, one can implement the manipulation, storage, and readout of the quantum information encoded in the different sublevels~\cite{Jelezko:2004,Fuchs:2009,Buckley:2010}. Furthermore, coherence times can be enhanced if one applies appropriate sequences of laser pulses and microwave fields~\cite{Lange:2010,Fuchs:2010,Naydenov:2011}, or transfers the quantum information from the electron spins to nearby nuclear spins by using the hyperfine interaction~\cite{Childress:2006,Jiang:2008,Fuchs:2011}. Such interactions also allow the implementation of few-qubit quantum registers~\cite{Dutt:2007,Neumann:2008}.

In addition, because NV centers couple to both optical and microwave fields, they can also be used as a quantum interface between optical and solid-state systems. This provides a promising platform to study novel quantum phenomena based on NV centers separated by long distances. For instance, quantum interference of two polarized optical photons produced by NV centers in two separate diamond samples~\cite{Sipahigil:2012} [or two separate NV centers in the same diamond sample~\cite{Bernien:2012}] and quantum entanglement between a polarized optical photon and a NV center qubit~\cite{Togan:2010} have recently been realized in experiment. 

Group-V endohedral fullerenes, consisting of a group-V atom (e.g., nitrogen) trapped inside a fullerene cage, are another spin system~\cite{Harneit:2002} that might be integrated in solid-state systems. Group-V endohedral fullerenes exhibit extraordinarily long spin relaxation times~\cite{Morton:2007} ($\sim1$ s at 4 K) due to the existence of the fullerene cage, which protects the enclosed spin from fluctuating perturbations in various host environments. The molecule N@C$_{60}$, which has electron spin $S=3/2$ coupled to the $^{14}$N nuclear spin $I=1$ via an isotropic hyperfine interaction, is a major member of the group-V endohedral fullerenes. By utilizing the technology of electron-spin resonance (or nuclear magnetic resonance), quantum operations and readout of the qubits encoded in the electron (or nuclear) spins of the nitrogen atoms of the molecule N@C$_{60}$ could be achieved.

Spin systems have similar or even longer coherence times than atomic systems. However, weak coupling to external fields leads to difficulties in the implementation of quantum gate operations. Solid-state systems, such as SC circuits, are an attractive platform for that purpose, as discussed in the next section.

\subsection{Superconducting qubits}\label{ssec:sc}

\begin{figure*}[tbp]
\includegraphics[width=6in]{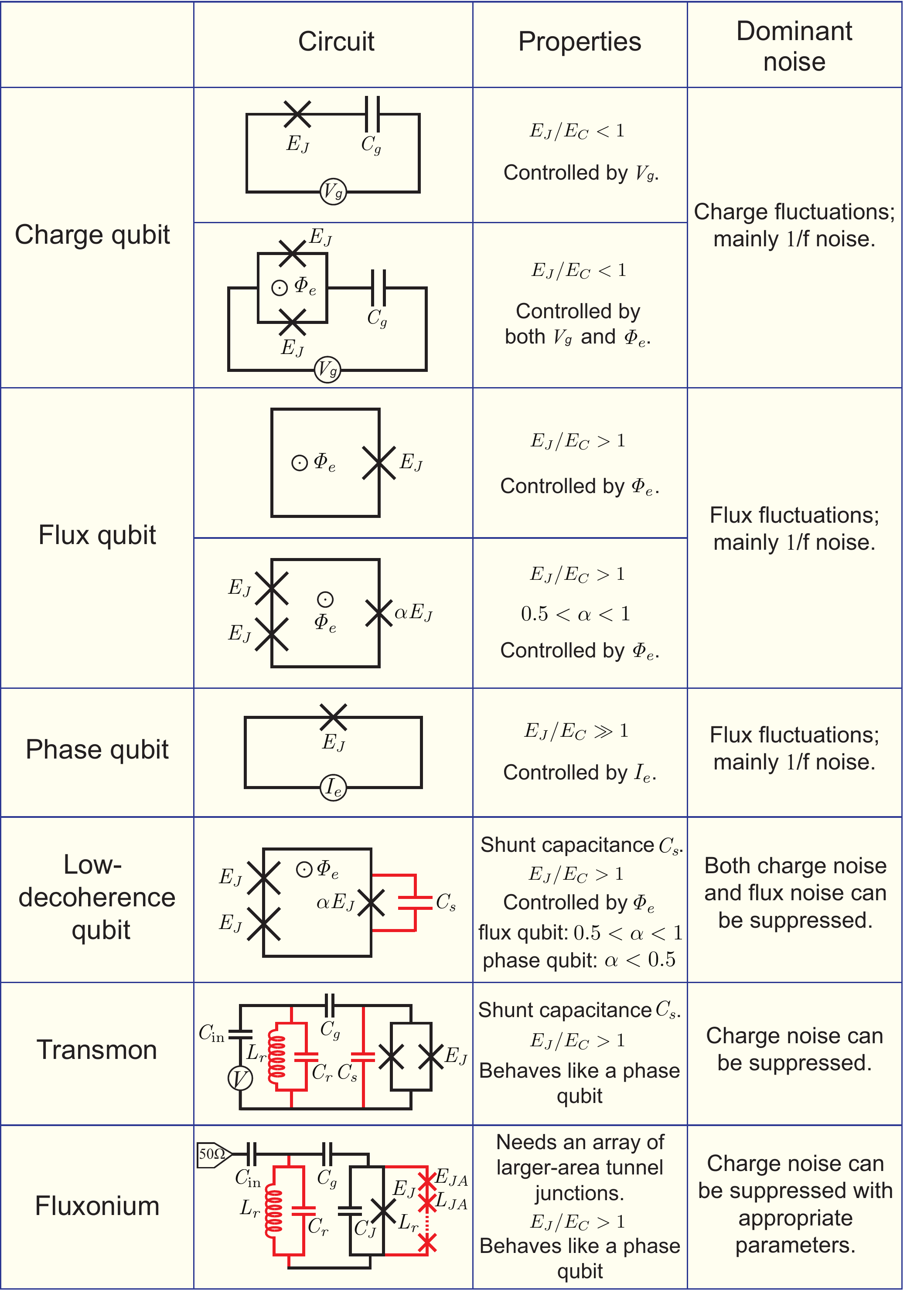}
\caption{(color online). Different types of superconducting qubits. The basic types are the first three ones from above. The bottom three can be thought of as improved versions, where additional components have been added.}
\label{SCQ}
\end{figure*}

SC qubit circuits based on Josephson junctions are macroscopic circuits, and they operate at temperatures of tens of mK. Although not microscopic in size, they can still behave quantum mechanically, allowing the observation of quantum coherence on a macroscopic scale. Compared to normal harmonic oscillators formed by {\it LC} circuits, in SC qubits the energy-level separation becomes nonuniform by introducing a nonlinearity via Josephson junctions. This property allows one to encode a qubit in the lowest two levels of a SC circuit for implementing quantum computing and simulation~\cite{Makhlin:2001,Burkard:2004,You:2005,Wendin:2007,Clarke:2008,Buluta:2009,You:2011,Georgescu:2011,Zagoskin:2011,Nation:2012}. These circuits have two important parameters: the Josephson coupling energy $E_J$ and the electrostatic Coulomb energy $E_C$. According to their topology and physical parameters, SC qubits can be divided into three kinds: charge qubits (using Cooper-pairs on a small island and $E_J/E_C<1$), flux qubits (using the circulating supercurrent states in a loop and $E_J/E_C>1$), and phase qubits (using the oscillatory states of the circuit and $E_J/E_C>1$), as shown in Fig.~\ref{SCQ}. These solid-state qubits can be controlled by the applied bias current, gate voltage, and microwave fields. All of these have been demonstrated in various experiments~\cite{Nakamura:1999,Wal:2000,Martinis:2002,Yu:2002}. Single-qubit and two-qubit gate times are $\sim1$ and $\sim$ 10--50 ns, respectively, while coherence times are currently $\sim 100~\mu$s and are growing steadily. Demonstrating the potential scalability of SC qubits, one experiment has integrated 512 qubits fabricated on a single chip~\cite{Harris:2012}. Thus, SC qubits are promising candidates for future quantum applications on a chip.

Superconducting qubits are sensitive to environmental noise from extrinsic and intrinsic decohering elements, which leads to short coherence times. Decoherence caused by extrinsic elements, such as the local electromagnetic environment, could be reduced using better design of the qubits and the surrounding circuitry, but the main intrinsic element that limits the coherence results from the hard-to-avoid low-frequency noise. For charge qubits, the dominant source of noise is charge fluctuators, such as trapped charges in the substrate and oxide layers of Josephson junctions. For flux and phase qubits, the noise from flux fluctuations dominates the decoherence. The development of more advanced SC qubit designs~\cite{You:2007,Koch:2007,Manucharyan:2009} recently ameliorated this problem.

For example, in three-junction flux qubits~\cite{Mooij:1999}, the effect of flux noise on the qubit can be reduced by decreasing $E_J$, but the charge noise can become increasingly important when decreasing $E_J$. The proposal by \textcite{You:2006} and \textcite{You:2007} improved the three-junction flux-qubit design by reducing the effective $E_J$ and adding a large shunt capacitor to the small junction, so as to reduce the effects of both charge and flux noise; see Fig.~\ref{SCQ}. Indeed, a recent experiment~\cite{Steffen:2010} demonstrated that this proposal is very effective in ameliorating the effect of low-frequency noise. A complementary proposal for a modified type of charge qubit, named transmon, was put forward by \textcite{Koch:2007}; see Fig.~\ref{SCQ}. The transmon qubit greatly reduces the charge dispersion, while the anharmonicity (which is necessary in order to prevent the qubit from turning into a harmonic oscillator) decreases by a much smaller amount. Because the qubit becomes less sensitive to charge variations, the need for electrostatic gates and tuning to a charge sweet spot becomes less necessary. In addition, the qubit in \textcite{Manucharyan:2009} is another improved SC qubit which was named fluxonium. In the fluxonium, a small junction is shunted by a series array of large-area tunnel junctions; see Fig.~\ref{SCQ}. By carefully choosing the parameters of the tunnel junctions, it is possible to protect the fluxonium from both charge and flux noise.

SC qubits can couple strongly to each other or to cavities via electromagnetic fields, which makes fast gate operations possible with current technology. However, strong coupling also leads to high sensitivity to noise and therefore short coherence times (several microseconds) compared to isolated atoms or spins. Thus, improving the coherence properties is a paramount priority for SC qubits. Recently, through dynamical decoupling with a flux qubit~\cite{Bylander:2011} or embedding the transmon qubit in a 3D circuit QED with a high $Q$ factor~\cite{Paik:2011,Rigetti:2012}, coherence times of SC qubits have been enhanced to around 10 $\mu$s.

\subsection{Cavities and resonators}

A cavity is one of the two basic elements of cavity QED~\cite{Scully:1997,Dutra:2005}. The quantized electromagnetic field in the cavity can interact with an atom (or spin or SC qubit), and exchange energy with it. Thus, a cavity can serve as a data bus in quantum information processing and transfer quantum data between different qubits. However, any real cavity system suffers from energy losses, which can be described by the quality factor $Q$. A higher $Q$ factor indicates a lower rate of energy loss in the cavity. In general, atoms and spins couple to conventional cavity systems, while SC qubits easily couple to SC resonators, such as SC coplanar waveguide (CPW) resonators and {\it LC} resonators, playing the role of cavities.

\subsubsection{Optical cavities}

\begin{figure}[t]
\includegraphics[width=3.4in]{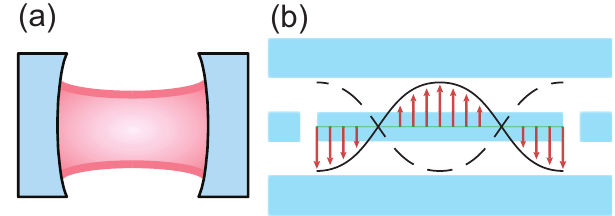}
\caption{(color online). Schematic diagram of (a) a cavity and (b) a coplanar waveguide (CPW) resonator.}
\label{CAV}
\end{figure}

Many types of cavity systems can be used to couple qubits to electromagnetic fields. The optical cavity consisting of two separated parallel mirrors is the most conventional cavity system, called the Fabry-Perot cavity, see Fig.~\ref{CAV}(a). In such a cavity, a standing-wave electromagnetic field can exist for a long time and interact with an atom (or spin) trapped in the cavity. With an appropriate design, a Fabry-Perot cavity can achieve high quality factors $Q\sim 3\times 10^8$~\cite{Hood:2001}, which provides a good platform for the realization of cavity QED. In other optical cavity systems, such as the microsphere cavity (dielectric spherically symmetric structure)~\cite{Vernooy:1998,Buck:2003}, the microtoroidal cavity (dielectric rotationally symmetric structure)~\cite{Armani:2003,Aoki:2006,Ozdemir:2011}, and the photonic band-gap cavity (periodic optical nanostructure that governs the motion of photons)~\cite{Lev:2004,Song:2005,Greentree:2006}, very high quality factors have also been achieved. The dynamics of a cavity can be described by the Hamiltonian:
\begin{equation}
H_{\rm cavity}\;=\;\sum_k\hbar\,\omega_k\left(a_k^{\dag}a_k\,+\,\frac{1}{2}\right),
\label{CARE}
\end{equation}
where $\omega_k$ is the frequency of the $k$th cavity mode, and $a_k^{\dag}$ ($a_k$) is the associated creation (annihilation) operator.

\subsubsection{Superconducting resonators}

Besides the above mentioned cavities, some other resonators, such as CPW and {\it LC} resonators, can also serve as cavities in SC circuits with low losses. These resonators can be described by the same Hamiltonian as that given in Eq.~(\ref{CARE}).

CPW resonators [see Fig.~\ref{CAV}(b)] have been realized in several experiments~\cite{Wallraff:2004,Sillanpaa:2007,Hofheinz:2008,Hofheinz:2009}. In these SC resonators, two ground planes are placed on the two sides of a central SC wire; this defines the CPW resonator (or transmission line resonator). Two gap capacitors play the role of the mirrors in a conventional optical cavity, and the distance between these capacitors defines the characteristic frequencies of the normal modes. Furthermore, the frequencies of the resonator can be adjusted by including a superconducting interference device (SQUID) array in the SC wire~\cite{Palacios:2008}. In general, the entire setup should be on the millimeter scale in order to fit the microwave frequency of the SC qubit, and it can be built by etching techniques. Recent experiments showed that gigahertz photons can make up to a million bounces before being lost in a high quality CPW resonator at low temperatures.

An {\it LC} resonator consists of an inductor and a capacitor with resonance frequency $\omega_r=1/\sqrt{LC}$. With appropriate design, the {\it LC} resonator can be integrated into a SC circuit and effectively serve as a cavity. Different from a CPW resonator, the {\it LC} resonator has only a single cavity mode.

One recent proposal for creating tunable cavities in SC circuits involves using a one-dimensional array of SC resonators as a medium and two SC qubits as tunable mirrors~\cite{Zhou:2008a,Zhou:2008b,Liao:2010}. The resonator array supports allowed energy bands for photon propagation, just as in a photonic crystal. When the frequency of a qubit that is coupled to one of the resonators is tuned to match the photon frequency, it can act like a perfectly reflecting mirror. Thus, by appropriately biasing two qubits coupled to two resonators in the array, one obtains an analog of a Fabry-Perot cavity.

\subsubsection{Nanomechanical resonators}

NAMRs have recently attracted considerable attention for their possible applications in future quantum technologies. With current experimental techniques, NAMRs can be built with quality factors $Q$ in the range of $10^3$ to $10^5$, and fundamental vibrational mode frequencies in the range from 10 MHz to 1 GHz, at low temperatures $T$~\cite{Cleland:1996,Knobel:2003,Lahaye:2004,Gaidarzhy:2005}. If the vibrational energy of the NAMR becomes larger than the thermal energy $k_BT$, then the mechanical oscillation can behave quantum mechanically. However, the observation of quantum behavior is challenging since it requires cooling the mechanical motion to extremely low temperatures and the ability to generate nonclassical states. Currently, many groups are devoting tremendous efforts to devising hybrid systems to create a coherent interface between NAMRs and other well-controlled quantum systems. Some focus on optomechanical systems involving an oscillating cantilever or an oscillating micromirror as a harmonic oscillator~\cite{Marquardt:2009}. Others focus on how to integrate NAMRs into SC circuits and couple them to each other~\cite{Armour:2002,Cleland:2004,Zhang:2005,Wei:2006,Sun:2006,Buks:2006,Xue:2007a,Xue:2007b,Xue:2007c,Etaki:2008,Wang:2008,Lahaye:2009,Xia:2009,Zhang:2009b,Pugnetti:2009,Liu:2010,O'Connell:2010,Teufel:2011,Shevchenko:2012}.

\section{Cavity quantum electrodynamics}\label{sec:cqed}

Cavity QED studies the interaction between matter and the field in a cavity. In the simplest case, a two-level system (TLS) and a quantized electromagnetic field can coherently exchange a quantum of energy (a single photon) back and forth in the cavity at the Rabi frequency, which is proportional to the system-field coupling strength. This energy exchange process is called Rabi oscillations~\cite{Scully:1997}. Unfortunately, in any real system, other undesirable losses, such as decay from the cavity (at rate $\kappa$) and from the atom (at rate $\gamma$), can take place. The coherent exchange of energy between the TLS and the cavity can be observed only when the Rabi frequency is much larger than the loss rates ($g\gg\kappa, \gamma$), which is known as the strong-coupling regime. Cavity QED can be realized in various systems, such as atom-cavity devices and spin-cavity systems, which have been studied for many years. Cavity QED with SC qubits has also been experimentally demonstrated. Table~II list data for different cavities and resonators.

\begin{table*}[t]
\label{Table2}
\begin{center}
\colorbox{mybo}{
\begin{tabular}{@{}p{4cm}|cccccc@{}}
\multicolumn{7}{l}{\rule[-3mm]{0mm}{8mm}{\bf TABLE II. ~Relevant parameters for different types of cavities in recent experiments.}}\\
\hline\hline
\rule[-3mm]{0mm}{8mm} {\bf Type} & \tabincell{c}{{\bf Frequency} \\ (Hz)} & \tabincell{c}{{\bf Temperature} \\(K)} & \tabincell{c}{{\bf Coupling} \\{\bf strength} \\$g/(2\pi)$ (MHz)} & \tabincell{c}{{\bf Coupling to} \\{\bf decay ratio} \\$g/\max(\kappa,\gamma)$} & \tabincell{c}{{\bf Cooperativity} \\$g^2/\kappa\gamma$} & \tabincell{c}{{\bf Quality factor} \\$Q$}\\
\hline
\tabincell{l}{{\bf Fabry-Perot cavity} \\\cite{Hood:2000}} & $1.9\times10^{14}$ & $3\times10^2$ & 110 & 7.7 & $\sim$ 30 & $>10^6$\\
\tabincell{l}{{\bf Microsphere cavity} \\\cite{Vernooy:1998}} & $3.5\times10^{14}$ & $\sim$ 10 & 20 & 2.9 & 22 & $1.5\times10^6$\\
\tabincell{l}{{\bf Microtoroidal cavity} \\\cite{Aoki:2006} \\\cite{Dayan:2008}} & $1.9\times10^{14}$ & $10^{-5}$ & 70 & 3.9 & $\sim 11$ & $\sim10^8$\\
\tabincell{l}{{\bf Photonic band-gap} \\{\bf cavity} \\\cite{Lev:2004}} & $3.5\times10^{14}$ & $3\times10^2$ & $1.7\times10^4$ & 3.9 & $2.5\times10^4$ & $4\times10^4$\\
\tabincell{l}{{\bf Coplanar waveguide}\\ {\bf resonator} \\\cite{Niemczyk:2010}} & $5.4\times10^9$ & $10^{-5}$ & $>~3\times10^2$ & $>~10^2$ & $>~10^4$ & $>~10^8$\\
\tabincell{l}{\textbf{\emph{LC}} {\bf resonator} \\\cite{Forn-Diaz:2010}} & $8.1\times10^9$ & $10^{-5}$ & 820 & $>$ 10 & $4\times10^4$ & $\sim 10^3$\\
\hline\hline
\end{tabular}}
\end{center}
\end{table*}

In general, the dynamics of a cavity-QED system consisting of a cavity mode and a TLS can be described by the universal Hamiltonian:
\begin{equation}
H \,=\, H_{\rm TLS} + H_{\rm cavity} + H_{\rm TLS-cavity}.
\end{equation}
Here $H_{\rm TLS}$ is the Hamiltonian of the TLS $H_{\rm TLS} = \hbar \omega_{\rm TLS} \sigma_{\rm TLS}^+\sigma_{\rm TLS}^-$, where $\omega_{\rm TLS}$ is the energy gap between the ground and excited states of the TLS, $\sigma_{\rm TLS}^+$ ($\sigma_{\rm TLS}^-$) denotes the TLS raising (lowering) operator. $H_{\rm cavity}$ is the Hamiltonian of the cavity mode $H_{\rm cavity} = \hbar\omega_{\rm cavity}\left(a^{\dag}a+1/2\right)$, where $a^{\dag}$ ($a$) is the creation (annihilation) operator of cavity photons with frequency $\omega_{{\rm cavity}}$. Lastly, $H_{\rm TLS-cavity}$ describes the interation between the TLS and the cavity mode:
\begin{equation}
H_{\rm TLS-cavity}\, =\, \hbar\, g_{\rm TLS-cavity} (\sigma_{\rm TLS}^+ +\sigma_{\rm TLS}^-)(a^{\dag}+a),
\end{equation}
where $g_{\rm TLS-cavity}$ is the coupling strength between the TLS and the cavity. 

In the interaction picture with respect to $H_0=H_{\rm TLS}+H_{\rm cavity}$, the Hamiltonian becomes
\begin{eqnarray}
H_I&\, =&\, \hbar \,g_{\rm TLS-cavity}\left(\sigma_{\rm TLS}^+ae^{-i\Delta t}+\sigma_{\rm TLS}^-a^{\dag}e^{i\Delta t}\right.\nonumber\\
&\, &\, \left.+\sigma_{\rm TLS}^+a^{\dag}e^{i\omega_+t}+\sigma_{\rm TLS}^-ae^{-i\omega_+t}\right),\label{NJC}
\end{eqnarray} 
where $\Delta=\omega_{\rm cavity}-\omega_{\rm TLS}$ is the detuning between the cavity mode and the TLS, and $\omega_+=\omega_{\rm cavity}+\omega_{\rm TLS}$. Depending on the relation between the system parameters, three different regimes are identified: the weak-coupling regime where $g\lesssim\kappa, \gamma$, the strong-coupling regime where $\kappa, \gamma\ll g\ll \omega_{\rm TLS}, \omega_{\rm cavity}$, and the ultrastrong-coupling regime where $g\sim \omega_{\rm TLS}, \omega_{\rm cavity}~(\gg \kappa,\gamma)$. In the weak- and strong-coupling regimes, the coupling strength $g_{\rm TLS-cavity}$ is much smaller than the frequencies of the two subsystems. At the same time, the so-called counter-rotating terms, i.e. those proportional to $\exp(\pm i\omega_+t)$, oscillate rapidly and their average over a time scale larger than $1/\omega_{\rm cavity}$ becomes zero. Thus, these terms lead to small and fast oscillations added to the otherwise smooth dynamics of the system, and they can be neglected in the rotating-wave approximation (RWA). The reduced interaction Hamiltonian becomes
\begin{equation}
H_I \,= \,\hbar g_{\rm TLS-cavity}(\sigma_{\rm TLS}^+ae^{-i\Delta t}+\sigma_{\rm TLS}^-a^{\dag}e^{i\Delta t}),
\end{equation}
or
\begin{eqnarray}
H&\, =&\, \hbar\omega_{\rm TLS}\sigma_{\rm TLS}^+\sigma_{\rm TLS}^-+\hbar\omega_c\left(a^{\dag}a+\frac{1}{2}\right) \nonumber\\
&\, &\, + \hbar g_{TLS-cavity}\left(a^{\dag}\sigma_{\rm TLS}^- + a\sigma_{\rm TLS}^+\right),
\label{JC}
\end{eqnarray}
which is known as the Jaynes-Cummings (JC) model. In the ultrastrong-coupling regime, where the coupling strength $g_{\rm TLS-cavity}$ is comparable with the energies of the TLS and the cavity mode, the counter-rotating terms also play an important role in determining the system properties and dynamics, and they cannot be neglected~\cite{Ashhab:2010}. The ultrastrong-coupling regime is an active research area.

\subsection{Atoms coupled to cavities}\label{ssec:ac}

In atom-cavity systems there are many significant physical phenomena that have been experimentally demonstrated in the strong-coupling regime, such as Rabi oscillations~\cite{Scully:1997}, which involve the exchange of photons between the atom and the cavity. The atom-cavity system can also be arranged as a waveguide to transfer or store quantum data~\cite{Kimble:2008,Zhou:2008a}. Because the coupling between the atom and the cavity mode is intrinsically proportional to a small constant $\alpha^{3/2}$, where $\alpha$ is the fine structure constant~\cite{Devoret:2007}, it is difficult to significantly increase the atom-cavity coupling strength.

In past years, rapid progress has been made in this field. Recent experimental demonstrations of strong coupling have been achieved in several kinds of cavities, such as FP cavities~\cite{Hood:2000,Hood:2001,Gleyzes:2007}, microsphere cavities~\cite{Vernooy:1998,Buck:2003}, photonic band-gap cavities~\cite{Lev:2004,Song:2005,Greentree:2006}, and microtoroidal cavities~\cite{Aoki:2006,Dayan:2008,Zhu:2010,Ozdemir:2011}. For example, in a toroidal cavity, the coupling between the atom and the cavity can exceed 700 MHz, and the relevant coupling-to-dissipation ratio can reach 40, which lies in the strong-coupling regime.

\subsection{Spins coupled to cavities}\label{ssec:sc}

Spins can also be placed in cavities and interact with cavity photons. Such a spin-cavity system, whose dynamics can usually be described by the same simple Hamiltonian (\ref{JC}) (where the index ``TLS'' is replaced by ``spin''), can therefore be employed to realize cavity QED. Furthermore, most spin systems, such as electron spins in quantum dots that are embedded in a solid-state chip and controlled by voltages, exhibit good scalability and tunability, which are vital elements for experimentally observing cavity-QED phenomena and designing quantum devices.

In the past few years, spin-cavity systems have been theoretically analyzed and experimentally demonstrated. Quantum dots can be integrated into micropillar cavities~\cite{Press:2007,Reithmaier:2004}, microdisk cavities~\cite{Imamoglu:1999,Peter:2005,Witzany:2011}, or photonic crystal cavities~\cite{Yoshie:2004,Hennessy:2007,Faraon:2008,Nomura:2010,Carter:2012}, and strong coupling with the photon field in the cavity can be achieved. Atomic impurity spins, especially NV centers, can also couple to microsphere cavities~\cite{Park:2006}, microtoroidal cavities~\cite{Chen:2011}, or photonic crystal cavities~\cite{Hanic:2006,Zagoskin:2007,Su:2009}.

\subsection{Superconducting qubits coupled to resonators}\label{ssec:scc}

\begin{figure*}[t]
\includegraphics[width=5.75in]{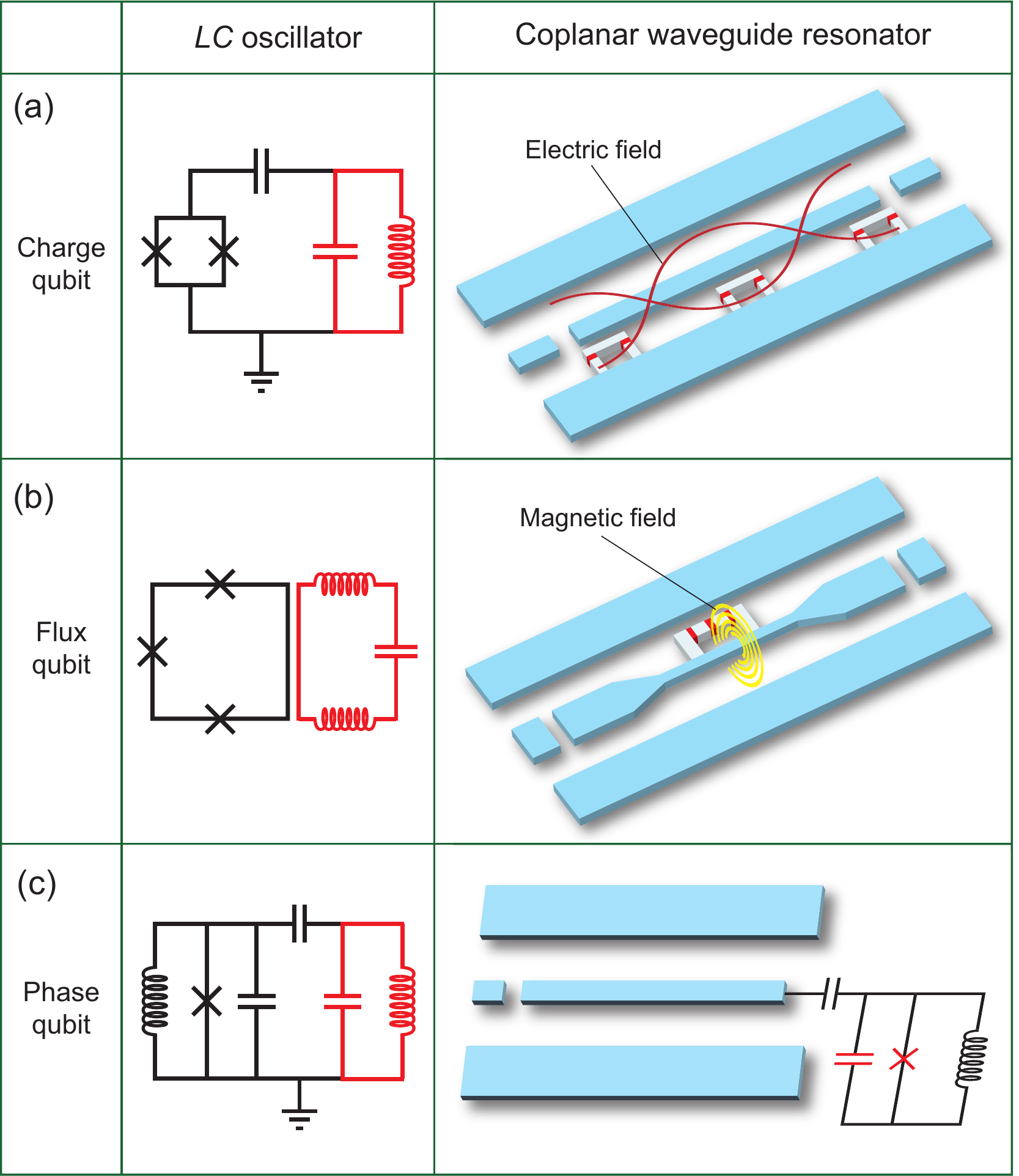}
\caption{(color online). Schematic diagrams of {\it LC} resonators (second column) and coplanar waveguide resonators (third column) coupled to three types of superconducting qubits.}
\label{SCC}
\end{figure*}

Recently, a growing new subfield in SC qubit research is finding physical phenomena in SC circuits analogous to the ones in atomic physics and quantum optics. A high quality microwave resonator can be coupled to SC qubits, which can be used to realize cavity-QED where the SC qubit is regarded as an artificial atom~\cite{Buluta:2011,You:2005,You:2011}. The resonator can be either a CPW resonator or an {\it LC} resonator. The dynamics of such SC circuits can be described by the Hamiltonian in Eq. (\ref{NJC}), where the index TLS is replaced by SC denoting  the SC qubit. In this Hamiltonian, the decay from the resonator and spontaneous emission are neglected.

The coupling strength between matter and the cavity mode is determined by both the transition dipole moment and the vacuum field strength. A SC qubit can have a large effective transition dipole moment, e.g., the effective electric dipole moment of a charge qubit is $\sim10^4$ times larger than that of an alkali atom~\cite{Blais:2004}. Moreover, in a quasi-1D  cavity, such as a CPW resonator, the microwave field is confined to a much smaller volume than in a conventional 3D optical or microwave cavity. This can make the field strength in the quasi-1D cavity much larger (about 100 times or more) than in a 3D cavity. Owing to the large dipole moment of a SC qubit and the strong electromagnetic field in a quasi-1D cavity, the SC qubit can couple to the quasi-1D cavity mode much more strongly than an atom or a spin couples to a 3D cavity. Therefore, it becomes easier to reach the strong, or even ultrastrong, coupling regime using a SC circuit consisting of a SC qubit and a quasi-1D cavity. Indeed, the strong~\cite{Wallraff:2004} and ultrastrong~\cite{Niemczyk:2010} coupling regimes have been experimentally realized in this SC-cavity system. In the usual strong-coupling regime, the counter-rotating terms (i.e., $\sigma_{\rm SC}^+a^{\dag}$ and $\sigma_{\rm SC}^-a$) can be neglected and the RWA is valid. However, in the ultrastrong-coupling regime~\cite{Ashhab:2010}, the counter-rotating terms also play an important role and cannot be neglected. Note that either an atom or a spin can also be placed in a quasi-1D cavity (see Sec.~\ref{ssec:hcqed}), but its coupling to the quasi-1D cavity mode is still much smaller than that of a SC qubit because the SC qubit has a much larger transition dipole moment.

For charge qubits, electric fields are well suited for coupling to the qubits~\cite{You:2003a}. Note that charge qubits can also be designed with a loop, such that they can also interact with magnetic fields~\cite{You:2003b}. A SC circuit involving a CPW resonator and a charge qubit was theoretically proposed by \textcite{Blais:2004,Blais:2007} and experimentally demonstrated by \textcite{Wallraff:2004}, where a strong electric coupling between a single photon and a charge qubit was achieved. In this setup, a charge qubit with two identical Josephson junctions is integrated into the ground planes of the transmission line at or near the antinode of the standing wave of the voltage on the SC wire for maximum coupling, as shown in Fig.~\ref{SCC}. The dynamics of this system can be described by the Hamiltonian in Eq.~(\ref{NJC}), and the strength of the coupling between the charge qubit and the SC resonator can in principle reach the ultrastrong-coupling regime~\cite{Devoret:2007}. A similar structure (Fig.~\ref{SCC}) and mechanism are also used for the electric coupling of phase qubits with SC resonators~\cite{Sillanpaa:2007,Hofheinz:2008,Hofheinz:2009}, where the phase qubits are placed on the two sides of the transmission line and coupled to it via capacitors, sitting on two antinodes of the electric field. The photon in the CPW resonator acts as a quantum bus that transfers quantum states between the two phase qubits.

Flux qubits can also couple to CPW resonators via the induced magnetic field~\cite{Yang:2003,Yang:2004,Peropadre:2010,Niemczyk:2010}, as shown in Fig.~\ref{SCC}. A flux qubit placed at or near an antinode of the standing wave of the current on the SC wire, can strongly couple to the SC resonator via the mutual inductance. In such a SC circuit, the vacuum Rabi splitting in the transmission spectrum was observed, which means that strong coupling was achieved. Furthermore, by placing an additional Josephson junction at the central SC wire, where the flux qubit is fabricated, the inductive coupling between the qubit and the resonator can be enhanced and can bring the system to the ultrastrong-coupling regime~\cite{Niemczyk:2010}.

The other type of resonators, {\it LC} resonators, can also be integrated into SC circuits and can couple to charge and phase qubits via capacitors (electric field) or flux qubits via the mutual inductance (magnetic field), see Fig.~\ref{SCC}. For example, in flux qubits, the lowest two quantum states, which have clockwise and anticlockwise supercurrents in the qubit loop, are used to denote the two basis states of the qubit. By employing the magnetic field produced by the current, the flux qubit can strongly couple to the {\it LC} circuit via a large mutual inductance between them. Such flux qubit-resonator systems have been experimentally demonstrated and vacuum Rabi oscillations have also been observed in experiment~\cite{Chiorescu:2004, Johansson:2006}, where a three-junction flux qubit is enclosed by a SQUID that is inductively coupled to the qubit. Recently, \textcite{Forn-Diaz:2010} observed the Bloch-Siegert shift in this flux qubit-resonator system and demonstrated that the coupling strength between them can lie in the ultrastrong-coupling regime.

\section{Theoretical principles and proposals for hybrid systems integrating atoms or spins in superconducting circuits}\label{sec:hqc}

\subsection{Atoms or spins coupled to superconducting resonators (without superconducting qubits)}\label{ssec:hcqed}

\begin{figure}[t]
\includegraphics[width=3.2in]{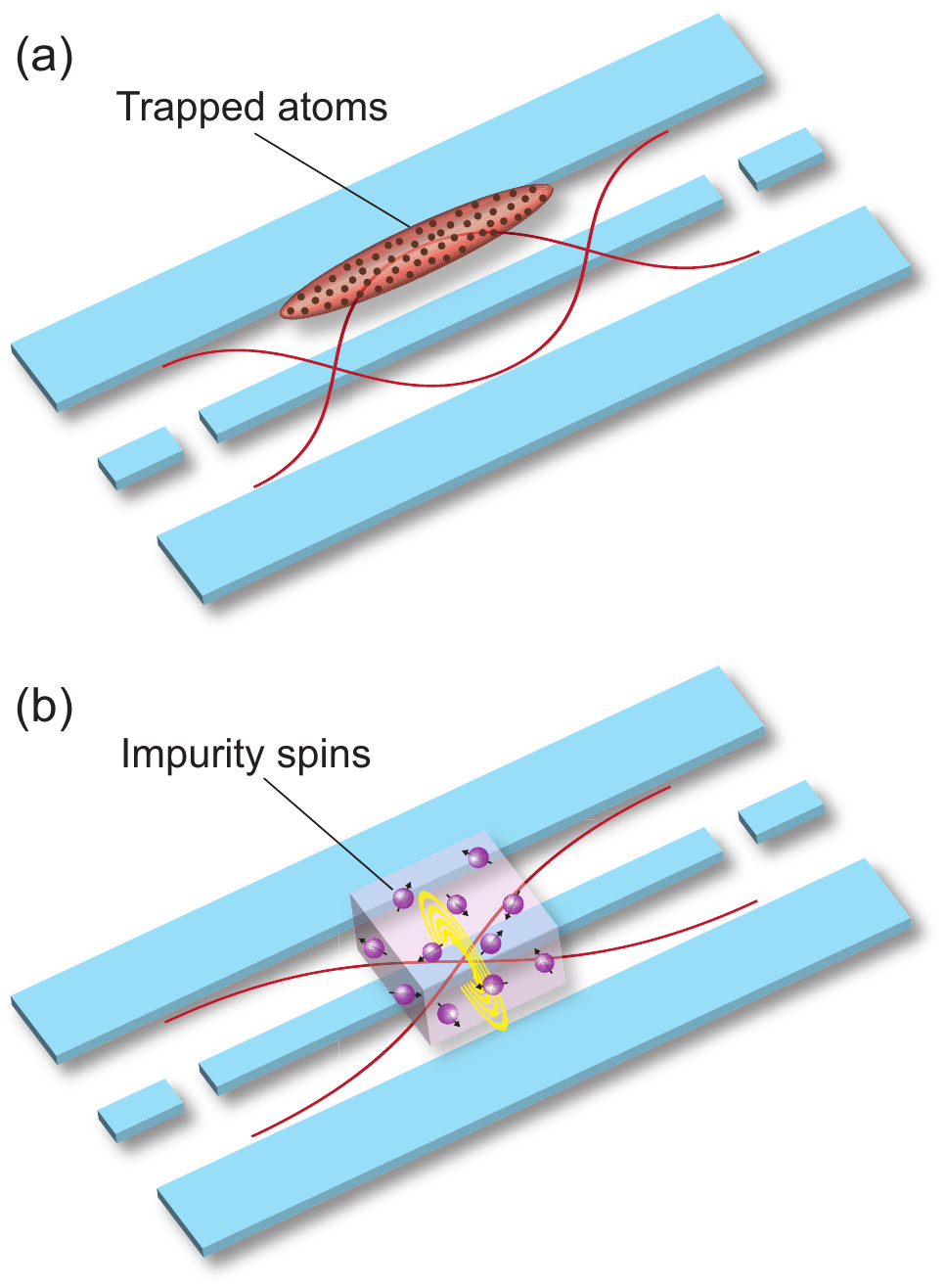}
\caption{(color online). Schematic diagrams of coplanar waveguide resonators with (a) atoms or (b) spins. The sinusoidal curves describe the electric field in the coplanar waveguide resonator and the yellow concentric circles in (b) denote the magnetic field.}
\label{AC}
\end{figure}

Besides SC qubits, atoms and spins can also couple to CPW resonators, producing a hybrid cavity-QED system. Atoms can be trapped either by electrostatic or magnetic fields generated by the chip or by trapping potentials from externally applied laser beams~\cite{Andre:2006,Rabl:2007}, and couple to a CPW resonator on the chip~\cite{Andre:2006,Rabl:2006,Rabl:2007,Tordrup:2008a,Tordrup:2008b,Zhang:2009a,Verdu:2009,Petrosyan:2009,Deng:2010}, see Fig.~\ref{AC}(a). 

Without extra trapping techniques, impurity spins, such as NV centers in diamond, can couple to a CPW resonator by placing the diamond sample on the resonator, or impurities can be directly created in the Si substrate~\cite{Imamoglu:2009,Wesenberg:2009,Kubo:2010,Schuster:2010,Bushev:2011,Amsuss:2011,Sandner:2012,Yang:2011a,Yang:2011b,Kubo:2012,Ping:2012,Ranjan:2012}, as shown in Fig.~\ref{AC}(b).

\begin{figure}[t]
\includegraphics[width=3.2in]{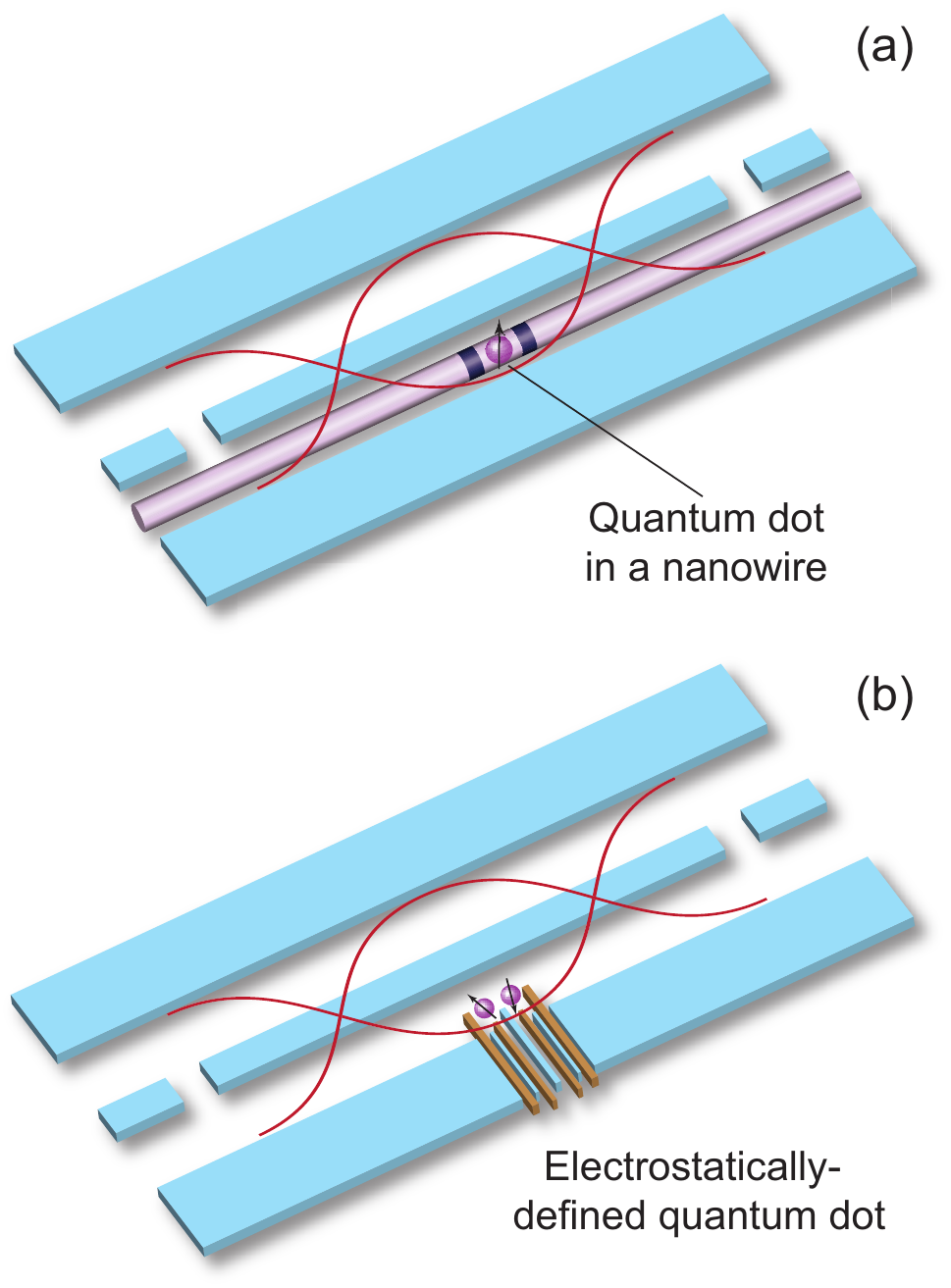}
\caption{(color online). Schematic diagram of coplanar waveguide resonators with spins in quantum dots. The sinusoidal curves describe the electric field in the coplanar waveguide resonator. (a) A nanowire quantum dot, and (b) an electrostatically defined quantum dot.}
\label{SC}
\end{figure}

In addition, quantum dots can also be integrated into a CPW resonator and coupled to the electromagnetic field in the resonator~\cite{Childress:2004}, as shown in Fig.~\ref{SC}. There have been a few recent proposals for coupling the spin state of the quantum dot to the electric field in the CPW resonator, generally mediated by the orbital state. In all of these proposals, the coupling between the field of the resonator and the orbital state is the electric-dipole coupling. The coupling between the spin state and the orbital state could be achieved via the application of an inhomogeneous magnetic field~\cite{Burkard:2006,Cottet:2010,Hu:2012}, spin-orbit coupling~\cite{Trif:2008,Hu:2012} or the exchange interaction~\cite{Jin:2012}. A number of recent experiments have demonstrated the coupling between quantum dots and CPW resonators.\textcite{Delbecq:2011} and \textcite{Frey:2012} demonstrated the coupling by using the dispersive frequency shift of the resonator as a probe for the charge state of the quantum dots. The results obtained using the resonator agreed with those obtained through transport measurements. In another experiment~\cite{Petersson:2012}, a spin qubit in a double quantum dot was manipulated via a classical microwave signal applied through the CPW resonator, with the spin-charge interface provided by the spin-orbit interaction. More recently, strong coupling between a CPW resonator and the charge degree of freedom of a double quantum dot was demonstrated through the observation of the vacuum Rabi splitting in spectroscopic measurements of the resonator~\cite{Toida:2012}. All of these experiments can be seen as first steps towards the coherent coupling between the spin state of quantum dots and a SC resonator in the few-photon regime.

\subsection{Atoms or spins coupled to superconducting resonators and qubits}\label{ssec:assrq}

\begin{figure*}[t]
\includegraphics[width=6.75in]{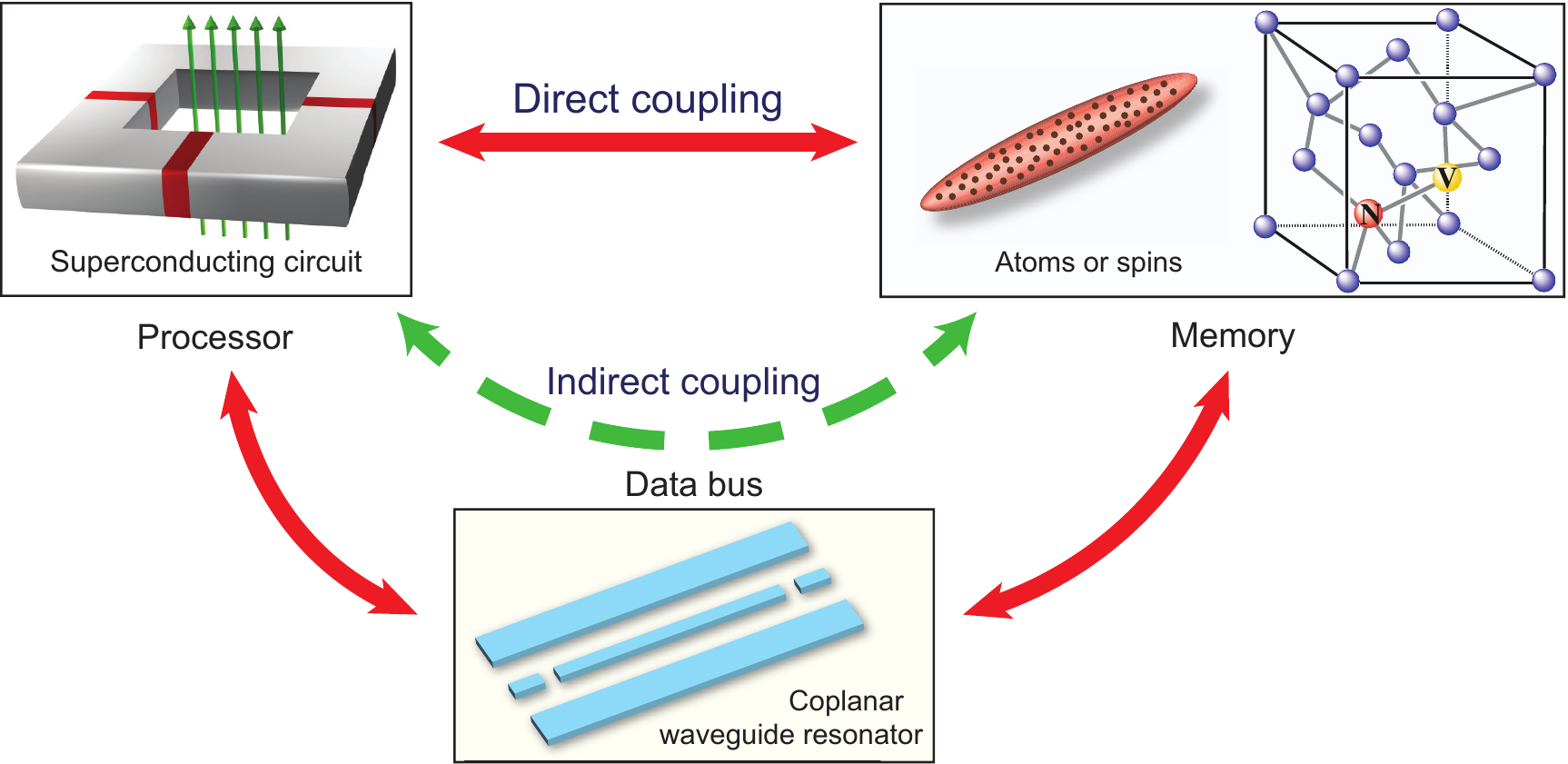}
\caption{(color online). Schematic diagram showing how to construct a hybrid quantum processor via a high-fidelity quantum data bus or ``bridge'' (indirect coupling) or without any intermediary (direct coupling). For fast and robust operations, superconducting circuits can serve as the processor; for long coherence times, atomic (or spin) systems can play the role of the memory in a hybrid quantum system. In the direct-coupling case, superconducting qubits couple with atoms (or spins) via electromagnetic fields. In the indirect-coupling case, a high-fidelity quantum resonator (e.g., coplanar waveguide resonator) acts as a data bus to transfer (quantum) information between the two components of the hybrid quantum system.}
\label{HS}
\end{figure*}

In an ideal HQC, the SC circuit provides the advantage of scalability on a small chip owing to the rapid progress on micro-lithography and micro-etching techniques, as well as ease of control due to the strong coupling of SC qubits with external fields. At the same time, atoms and spins can be integrated into the circuit by using trapping or doping techniques. Such a HQC involving SC qubits and atoms (or spins) thereby combines ``the best of two worlds'': the rapid operations of the SC circuits and the long coherence times of the atoms (or spins), as well as scalability, see Fig.~\ref{HS}.

However, effectively integrating such systems and controlling the resultant hybrid circuits are still a challenge. In the remainder of this section, we first introduce the basic mechanisms of HQCs, and then highlight recent theoretical proposals and experimental demonstrations for implementing various types of HQCs.

Atoms and spins can be initialized, manipulated and measured through their interaction with electromagnetic fields. SC qubits also interact with electromagnetic fields, which are used to initialize, manipulate and measure them. Therefore, electromagnetic fields can be utilized to couple atomic (or spin) systems to SC qubits in order to facilitate the transfer of information between the two systems.

There are two different types of coupling to electromagnetic fields: electric coupling and magnetic coupling. Generally, most atomic systems (including ions and molecules) couple to photons via the electric field, while spins involve magnetic coupling. Moreover, electric fields are well suited for coupling to SC charge and phase qubits, while flux qubits couple more easily to magnetic fields threading the qubit loop. Note that charge qubits can also be designed with a loop, such that they also interact with magnetic fields.

Atoms (or spins) and SC qubits can interact with each other directly through electric or magnetic fields. Alternatively, they can be linked indirectly via a quantum ``bridge'' or data bus, which mediates the exchange of quantum information between the atomic (or spin) memory and the SC processor.

Each type of HQC has its advantages and disadvantages. Direct-coupling hybrid circuits have simple and minimal structures. However, the direct coupling between an atom (or spin) and a SC qubit is usually weak and not tunable. In such cases, indirect coupling through a data bus, which interacts strongly with both systems, can be advantageous. As an additional advantage of indirect coupling, quantum cavities, such as CPW resonators, can be much larger in size than SC qubits, so that it is relatively easy to place many qubits in the same cavity.

\subsubsection{Direct coupling}

In this case, atomic (or spin) systems couple directly to SC qubits via electromagnetic fields. The dynamics of this type of hybrid circuit can be described by a total Hamiltonian consisting of terms that describe the two components separately and the interaction $H_{\rm int}$ between them~\cite{Imamoglu:1999},
\begin{eqnarray}
&& H_{\rm atom}\,+\, H_{\rm SC}\, +\, H_{\rm int}\nonumber\\
&& ~~~~=\hbar\,\omega_{\rm atom}\,\sigma^+_{\rm atom}\,\sigma^-_{\rm atom}\,+\,\hbar\,\omega_{\rm SC}\,\sigma^+_{\rm SC}\,\sigma^-_{\rm SC} \nonumber\\
&& ~~~~~~~ +\,\hbar\,g_{\rm atom-sc}\,(\sigma_{\rm atom}^+\,\sigma_{\rm SC}^- \,+\, \sigma_{\rm atom}^-\,\sigma_{\rm SC}^+),
\label{ATSC}
\end{eqnarray}
where $g_{\rm atom-sc}$ is the coupling strength between the atom and the SC qubit. The interaction term describes the dynamics of the energy exchange between the two systems. Generally, this term can be ignored when the two coupled systems are far detuned from each other,
\begin{equation}
|\,\omega_{\rm atom}\,-\,\omega_{\rm SC}\,|\,\gg\, g,\nonumber
\end{equation}
and is active only for resonant or near-resonant systems,
\begin{equation}
|\,\omega_{\rm atom}\,-\,\omega_{\rm SC}\,|\,\ll\, g.\nonumber
\end{equation}
Note that the index ``atom'' is replaced by ``spin'' for the hybrid circuit consisting of a spin and a SC qubit.

\subsubsection{Indirect coupling}

A high quality quantum cavity, such as an {\it LC} resonator or a CPW resonator, can be employed as an intermediary to link atoms (or spins) and SC qubits. Because of recent experimental advances on CPW resonators, much recent attention has focused on this type of resonator.

SC qubits can be integrated on the CPW resonator and couple to the electric or magnetic field of the resonator. A charge or phase qubit placed at or near an antinode of the standing wave of the voltage on the SC wire can couple strongly to the electric field of the standing wave~\cite{You:2003a,Blais:2004,Schoelkopf:2008}. A flux qubit, on the other hand, couples more naturally to the magnetic field and would therefore be placed at or near an antinode of the standing wave of the current~\cite{Niemczyk:2010}.

In a frame rotating at the cavity frequency, the dynamics of this system can be described by the Hamiltonian:
\begin{eqnarray}
&&H_{\rm SC}\, +\, H_{\rm sc-cavity}\nonumber\\
&& ~~~~= \frac{\hbar\,\delta_{\rm SC}}{2}\,\sigma^z_{\rm SC}\,+\,\hbar\,g_{\rm sc-cavity}(\sigma_{\rm SC}^+\,a \,+\, \sigma_{\rm SC}^-\,a^{\dag}),
\label{SCRE}
\end{eqnarray}
where $\delta_{\rm SC}$ is the detuning of the SC qubit from the resonance frequency of the cavity, and $g_{\rm sc-cavity}$ denotes the coupling strength between the SC qubit and the cavity field.

Atoms (or spins) can also be integrated on the CPW resonator and couple to the electromagnetic field~\cite{Andre:2006,Rabl:2006,Rabl:2007,Tordrup:2008a,Tordrup:2008b,Verdu:2009,Petrosyan:2009}, as described in Sec.~\ref{ssec:hcqed}. The dynamics of this system, consisting of a single atom (or spin) and a cavity, can be described by the Hamiltonian:
\begin{eqnarray}
& & H_{\rm atom}\,+\,H_{\rm atom-cavity}\nonumber\\
& & ~~~~ = \hbar\,\delta_{\rm atom}\,\sigma_{\rm atom}^+\,\sigma_{\rm atom}^- \nonumber\\
& & ~~~~~~~ +\,\hbar\,g_{\rm atom-cavity}\left(\sigma_{\rm atom}^+\,a\,+\,\sigma_{\rm atom}^-\,a^{\dag}\right),
\label{ATRE}
\end{eqnarray}
where $\delta_{\rm atom}$ denotes the atom-cavity detuning, and $g_{\rm atom-cavity}$ is the coupling strength between the single atom and the field. For a spin-cavity system, the index ``atom'' in Eq.~(\ref{ATRE}) is replaced by ``spin''. If an ensemble with $N$ atoms (or spins) is utilized, the Hamiltonian takes the Tavis-Cummings form,
\begin{eqnarray}
& & H_{\rm atom}\,+\,H_{\rm atom-cavity}\nonumber\\
& & ~~~~ = \hbar\,\delta_{\rm atom}\,\pi_{\rm atom}^{\dag}\,\pi_{\rm atom}\nonumber\\
& & ~~~~~~~ +\,\hbar\,g_{\rm atom-cavity}^{\prime}\left(\pi_{\rm atom}^{\dag}\,a\,+\,\pi_{\rm atom}\,a^{\dag}\right),
\label{TC}
\end{eqnarray}
where $\pi_{\rm atom}=(1/\sqrt{N})\sum_i\sigma_{{\rm atom},i}^-$ [$\pi_{\rm atom}^{\dag}=(1/\sqrt{N})\sum_i\sigma_{{\rm atom},i}^+$] is the collective atomic annihilation (excitation) operator, and the index $i$ denotes the different atoms (or spins) in the ensemble. Here the effective coupling strength becomes $g_{\rm atom(spin)-cavity}^{\prime}=\sqrt{N}g_{\rm atom(spin)-cavity}$~\cite{Dicke:1954}, which helps achieve strong coupling between the atoms (or spins) and the cavity. It should be noted, however, that the effective coupling strength cannot be increased indefinitely by increasing the size of the cavity and along with it the number of atoms (or spins) that can fit inside the cavity. If the size of the cavity is increased, the coupling strength per atom (or spin) decreases as the inverse of the square-root of the mode volume, such that the gain in increasing number is counterbalanced by the reduction in coupling strength per atom (or spin).

There are two main protocols for transferring quantum information through the intermediary bridge: via the exchange of either real or virtual photons. In the case of real-photon-based protocols, when two subsystems are coupled to a fixed cavity, one subsystem can be tuned into resonance with the cavity for a period of time to transfer the quantum information from the subsystem to the cavity and then the other subsystem is tuned into resonance with the cavity for a period of time, so as to finally transfer the quantum information into this latter subsystem. Alternatively, if the cavity is tunable, it can be tuned into resonance with one of the two subsystems for a period of time such that quantum information is transferred from the subsystem to the cavity, and then the cavity is tuned into resonance with the other subsystem for another period of time to complete the information transfer. In the case of virtual-photon-mediated interactions, the atoms (or spins) and the SC qubit are tuned into resonance with each other, while the cavity is off-resonance with them and typically has a higher frequency. The cavity can then be adiabatically eliminated from the physical picture, leading to the effective interaction Hamiltonian~\cite{Frohlich:1950,Nakajima:1955,Imamoglu:1999}
\begin{equation}
H_{\rm eff}= \hbar\, g_{\rm eff}\,(\sigma_{\rm atom}^+\,\sigma_{\rm SC}^-\,+\,\sigma_{\rm atom}^-\,\sigma_{\rm SC}^+).
\label{EFF}
\end{equation}
The effective coupling strength is
\begin{equation}
g_{\rm eff}=\frac{g_{\rm atom-cavity}\,g_{\rm sc-cavity}}{\Delta},\nonumber
\end{equation}
where $\Delta$ is the detuning of the two systems from the cavity frequency.

In order to achieve much longer coherence times in spin ensembles, the information could further be transferred from electron spins to nuclear spins by using hyperfine interactions between them~\cite{Amsuss:2011,Wu:2010,Childress:2006,Dutt:2007,Jiang:2008,Sar:2012}. This would improve the coherence times from several hundred microseconds in the electron spins to seconds in the nuclear spins, which would be a great improvement.

\subsection{Direct-coupling hybrid circuits}\label{ssec:dcc}

\begin{figure}[t]
\includegraphics[width=3.2in]{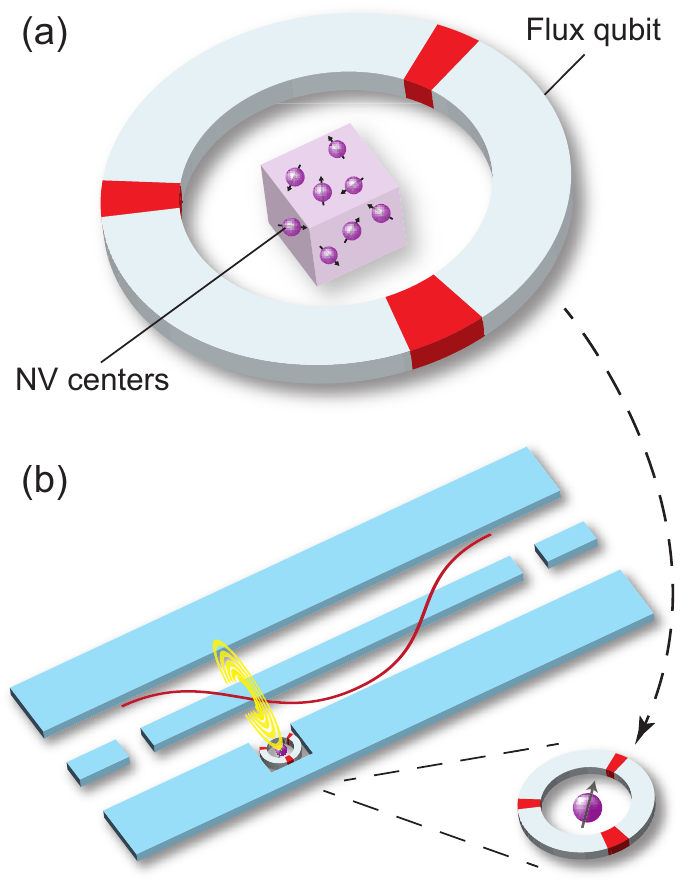}
\caption{(color online). (a) Schematic diagram of a hybrid system consisting of  a spin ensemble and a superconducting flux qubit. (b) Schematic diagram of a hybrid system integrating the unit in (a) with a superconducting resonator.}
\label{FLNV}
\end{figure}

Recently, \textcite{Marcos:2010} proposed a hybrid SC circuit with a three-junction flux qubit magnetically coupled to a single NV center or an ensemble of NV centers in diamond located at the center of the SC loop, as shown in Fig.~\ref{FLNV}(a). These two systems have similar energy splittings: the energy gap between the two eigenstates of the flux qubit is typically a few GHz, while the NV centers' electronic ground state has a zero-field splitting $\Delta \sim 2\pi\times2.87$ GHz between the $m_s=0$ and $m_s=\pm1$ sublevels. By introducing an external magnetic field $W^{\rm ext}$, the NV center can be tuned into resonance with the flux qubit, and the flux qubit can be brought near the degeneracy point of the clockwise and anticlockwise supercurrent states, which produce an additional magnetic field $W^{\rm FQ}$. The interaction between the NV centers and the magnetic field produced by the flux qubit leads to a coupling between the two systems.

The details can be described as follows. The axis of the NV centers is defined as the $z$-axis and it can be taken to lie in the plane of the flux qubit. The component of the external field that is parallel to the $z$-axis isolates the NV center as a two-level subsystem involving the states $m_s=0$ and $m_s=+1$, while the component of the field that is perpendicular to the qubit loop is set to half a flux quantum and brings the flux qubit near the degeneracy point. The dynamics of this process can be described by the Hamiltonian:
\begin{equation}
H=\frac{\varepsilon}{2}\sigma_z + \lambda\sigma_x+DS_z^2+W_z^{\rm ext}S_z+\sigma_z\vec{W}^{\rm FQ}\cdot\vec{S},
\label{NVFL}
\end{equation}
where $\vec{\sigma}$ denotes the Pauli operators of the flux qubit, $\vec{S}$ describes the spin of the NV center, $\lambda$ is the tunneling strength between the two supercurrents states, and $\varepsilon$ is the bias in the two-well limit of the flux qubit, which is controlled by the external field perpendicular to the qubit loop. $D=2.87$ GHz is the zero-field splitting of the NV center. $W_{z}^{\rm ext}=g_e\mu_BB_z$ is the parallel component of the external magnetic field, which adjusts the energy splitting of the NV center and $\vec{W}^{\rm FQ}$ corresponds to the induced magnetic field of the two currents of the flux qubit.

By rotating the flux-qubit terms by an angle $\cos\theta\equiv \epsilon/2\omega$ (where $\omega\equiv\sqrt{\varepsilon^2/4+\lambda^2}$) via a unitary transformation and making the RWA, the effective Hamiltonian (in a frame rotating with frequency $\omega$) of near-resonance interaction between the NV center and the flux qubit can be obtained from Eq.~(\ref{NVFL}):
\begin{equation}
H_{\rm eff}=\frac{\delta}{2}\sigma_z^s + \frac{\cos\theta}{2}W_z^{\rm FQ}\sigma_z\sigma_z^s+\left(\frac{\sin\theta}{\sqrt{2}}W_{\bot}^{\rm FQ}\sigma_-\sigma_+^s+H.c.\right),
\label{NVFLE}
\end{equation}
where $\vec{\sigma}^s$ denotes the Pauli operators of the electron-spin states of the NV center, $\vec{\sigma}$ refers to the Pauli operators of the the flux qubit in the rotated basis describing the two eigenstates of the flux qubit, and $\delta=D+W_z^{\rm ext}-\omega$ is the detuning of the NV center from the eigenenergy of the flux qubit. The last two terms of this effective Hamiltonian describe the exchange of energy between the NV center and the flux qubit.

In order to obtain the maximum $g_{\rm max}~(=W^{\rm FQ}_{\bot}/\sqrt{2})$ of the coupling strength $g=\sin\theta\,W_{\bot}^{\rm FQ}/\sqrt{2}$, one can bias the flux qubit at the degeneracy point $\varepsilon=0$, where $\theta=\pi/2$. Simultaneously, at this point the energy levels and dynamics are insensitive to small fluctuations of $\varepsilon$.

With a flux qubit of size $L \sim 1~\mu$m and critical current $I_C \sim 0.5~\mu$A, the coupling strength between the NV center and the flux qubit is $g/2\pi\sim10$ kHz~\cite{Marcos:2010}. However, the effective coupling between the flux qubit and a single NV center is too weak for any realistic demonstration, using current technology. This coupling strength can be enhanced by reducing the size of the flux qubit or replacing the single NV center with an ensemble~\cite{Marcos:2010}. Recently, this proposal was demonstrated in experiment~\cite{Zhu:2011} [introduced in Sec.~\ref{ssec:dcnv}]. However, because the size of the flux qubit limits the number of NV centers in the ensemble, the coupling strength cannot be substantially enhanced in this way.

This circuit can also achieve the coherent coupling of two or more NV centers by employing the flux qubit as a virtual intermediary, which is a possible protocol to eventually implement many-qubit quantum gate operations for spins. In addition, if one could integrate this circuit encompassing a single NV center in a CPW resonator, as shown in Fig.~\ref{FLNV}(b), the single NV center could be effectively coupled to the photon via the magnetic field in the resonator~\cite{Twamley:2010}. Furthermore, this proposal can also be used to couple an atom ensemble and a flux qubit. For example, \textcite{Hoffman:2011} has discussed a scheme to couple trapped $^{87}$Rb atoms to a flux qubit via magnetic-dipole coupling, where the atoms are trapped by an ultrathin fiber to less than 10 $\mu$m above the surface of the flux qubit. Also, based on the strong coupling between the spin ensemble and the flux qubit and the tunable coupling between nearest-neighbor flux qubits, one can construct a hybrid array with flux qubits and NV centers to simulate a Jaynes-Cummings lattice~\cite{Hummer:2012}, which can be used to demonstrate the transition between localized and delocalized phases.

\subsection{Indirect-coupling hybrid circuits}\label{ssec:icc}

In indirect-coupling HQCs, a quantum cavity (e.g. CPW resonator) transfers the quantum information between atoms (or spins) and SC qubits. As indicated in Sec.~\ref{ssec:scc}, the CPW resonator can be strongly coupled to the SC qubits. In order to achieve strong coupling between the atoms (or spins) and the CPW resonator, a large number of atoms (or spins) can be placed on the resonator, which is typically much larger than a SC qubit.

\begin{figure}[t]
\includegraphics[width=3.2in]{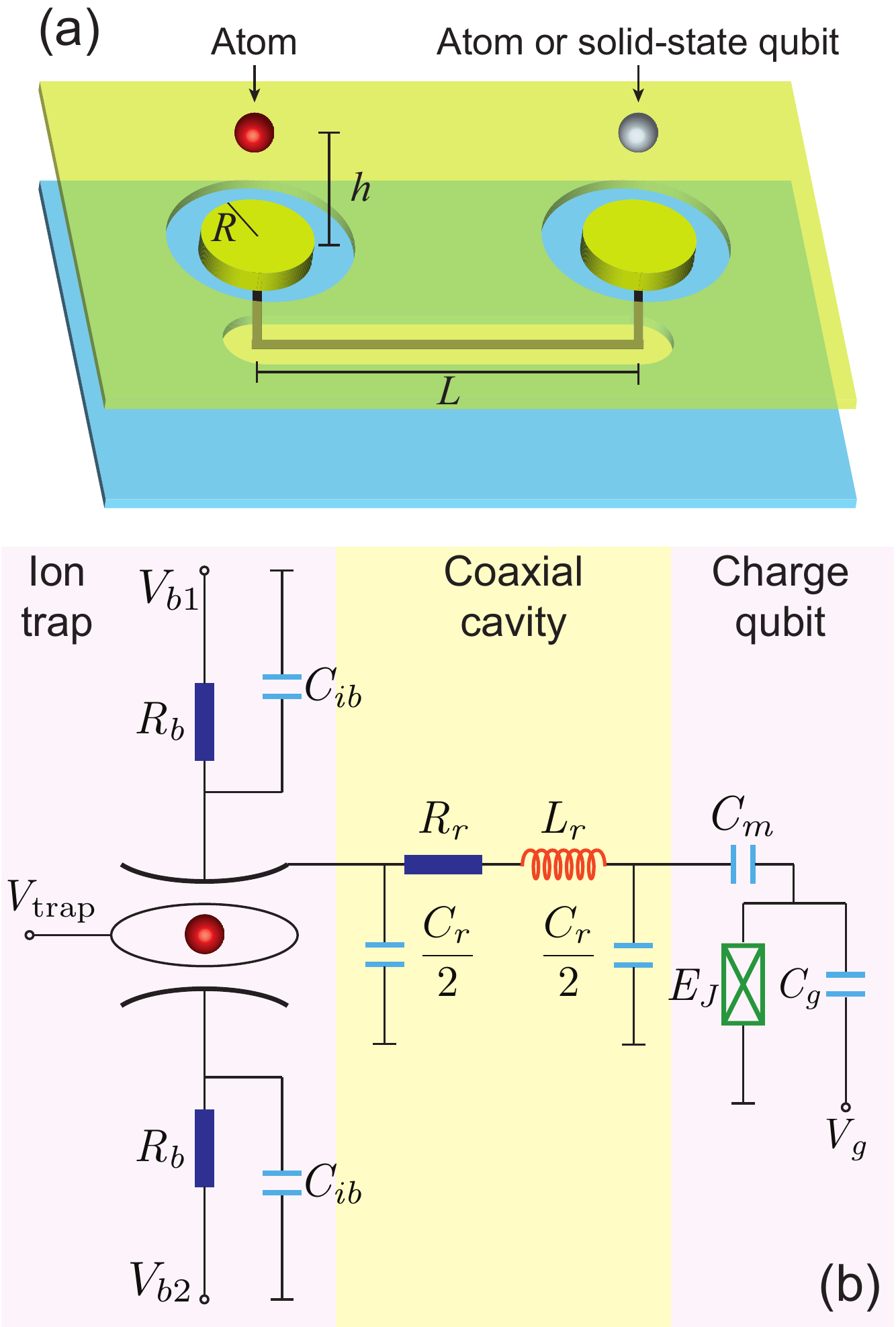}
\caption{(color online). Two early proposed architectures for implementing hybrid systems. (a) \textcite{Sorensen:2004} proposed a design with a superconducting wire in a cavity to link two systems (e.g., a trapped atom and a superconducting qubit), and (b) \textcite{Tian:2004} proposed a scheme using a coaxial cavity to link an atom and a superconducting qubit. (a) and (b) are adapted from \textcite{Sorensen:2004} and \textcite{Tian:2004}, respectively.}
\label{EHS}
\end{figure}

\subsubsection{Atomic hybrid quantum circuits}\label{sssec:ahqc}

Atomic systems can be integrated into CPW resonators and implement quantum information processing with SC circuits. This kind of hybrid circuit could be called atomic hybrid quantum circuits. Owing to recent progress in cooling and trapping techniques, many groups are devoting efforts to developing such circuits~\cite{Sorensen:2004,Tian:2004,Rabl:2006,Verdu:2009,Petrosyan:2009,Tordrup:2008a,Tordrup:2008b,Zhang:2009a,Deng:2010}.

\bigskip

\emph{Early proposals for hybrid systems ---}
Before focus was placed on CPW resonators, some other couplers or bridges were used in several early hybrid-circuit proposals. \textcite{Sorensen:2004} theoretically designed a hybrid system consisting of two individual atoms (or one atom and one solid-state qubit) that are linked by a SC wire, as shown in Fig.~\ref{EHS}(a). A single atom, used as a qubit, is trapped above a conducting disk, which is connected to a second disk via a thin SC wire. Because of the capacitive coupling between the atom and the conductor, a single and long-lived mode exciton can be produced in the conductor, which can be transferred to the SC cavity via the SC wire. On the other side of this setup, another atom or a solid-state qubit (such as a SC qubit) is capacitively coupled to the second disk connected to the SC wire. The dynamics of this setup can be described as follows. If the atom is excited to a Rydberg state with a large dipole moment, the charge distribution in the SC wire is modified accordingly. This charge redistribution extends throughout the wire. Another atom or a solid-state qubit, placed at the other end of the wire would be affected by the change in the electric field, which is produced by the charge in the conductor below it. Consequently, these two atoms (or an atom and a solid-state qubit) are effectively coupled through their mutual interaction with the SC wire. This coupling can be either electrostatic or electrodynamic. In such a hybrid system, the coupling between the atom and the SC wire could theoretically exceed 1 MHz, which is much larger than the decoherence rates of Rydberg atoms and solid-state qubits. In spite of its simplicity, experimental difficulties, such as the trapping of atoms at exact positions, are still a challenge, and this
proposal has not yet been realized.

Another early prototype of atomic HQCs proposed by \textcite{Tian:2004} used a SC cavity made of two parallel cylindrical rods to link a trapped ion and a SC charge qubit. This proposed circuit consists of three parts: the quantum optical side (a trapped ion), the bridge (a coaxial cavity), and the solid-state side (a charge qubit), as show in Fig.~\ref{EHS}(b). On the quantum optical side, a charged ion is trapped in a 1D harmonic potential. Two atomic ground states, which are coupled by a laser-induced Raman transition, are used to represent the two qubit states, and they can be controlled by the two trap electrodes via the electric field [see Fig.\ref{EHS}(b)]. On the solid-state side, a charge qubit is utilized in this proposal, and quantum operations on it can be implemented by adjusting the gate voltage. Between the ion and the solid-state qubit, a superconducting cavity plays the role of the bridge: one side is capacitively coupled to one of the trap electrodes of the optical part, and the other side interacts with the charge qubit via another contact capacitor. Consequently, an effective coupling between the ion and the charge qubit can be derived by adiabatically eliminating the cavity. By employing a ``fast swap'' gate that can be implemented on a timescale of nanoseconds~\cite{Tian:2004} and together with single-qubit rotations, one could achieve entanglement and information exchange between the ion and charge qubit. Also, this proposal has not yet been realized in experiment.

\bigskip

\begin{figure}[t]
\includegraphics[width=3.2in]{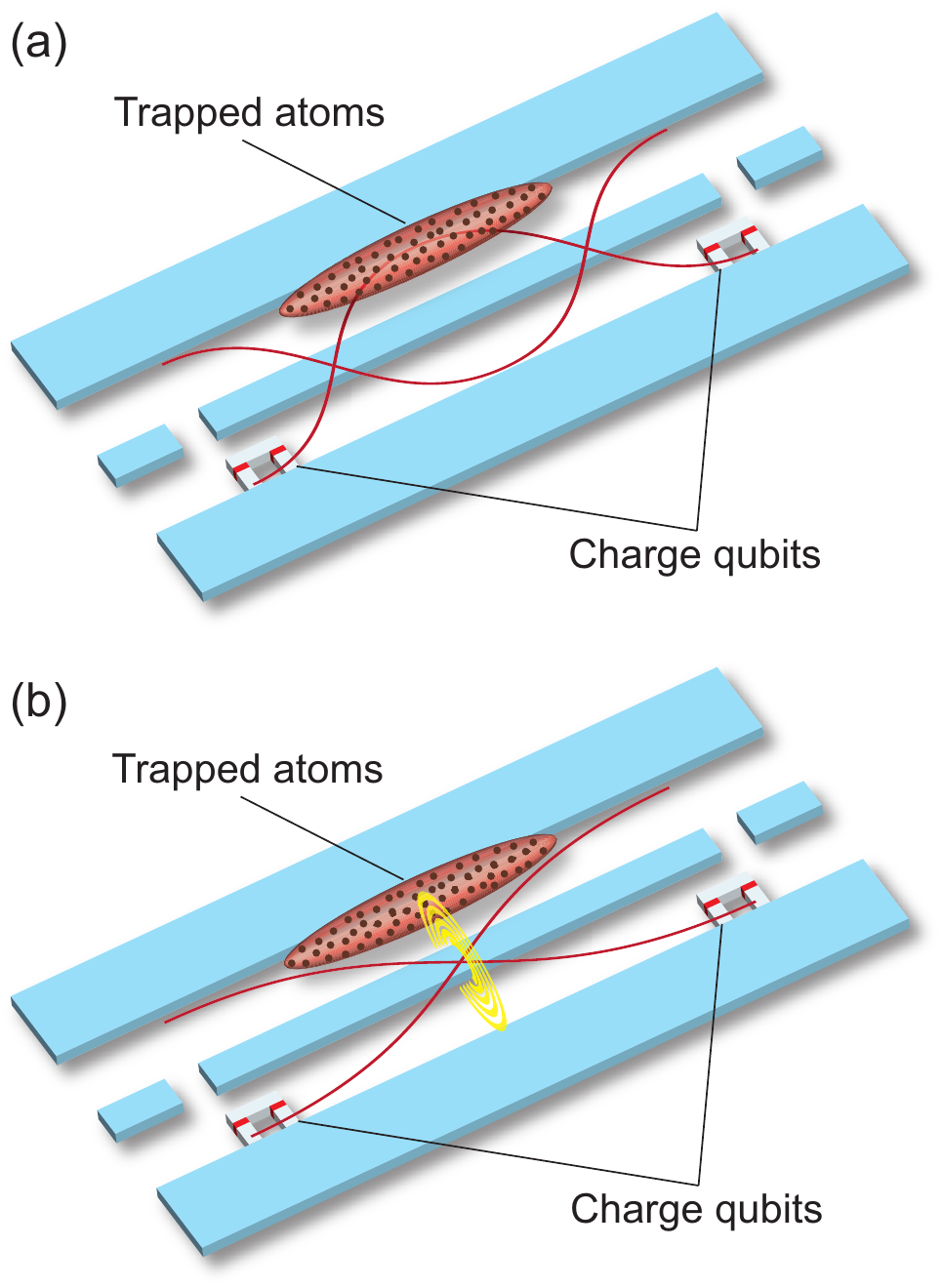}
\caption{(color online). Schematic diagrams of two different types of hybrid systems consisting of atomic ensembles and superconducting resonators classified according to the nature of coupling between the atoms and the resonator. (a) The ensemble of atoms couples to the coplanar waveguide resonator via the {\it electric} field. (b) The ensemble of atoms couples to the coplanar waveguide resonator via the {\it magnetic} field. In both (a) and (b), superconducting qubits are also coupled to the coplanar waveguide resonator. Thus, this resonator can be utilized as a data bus to indirectly couple atoms and SC qubits. Here the charge qubits are integrated on the ground plane at the antinode of the electric field, whereas flux qubits are integrated in the resonator at the antinode of the magnetic field.}
\label{HSAS}
\end{figure}

\emph{Electric-coupling HQCs ---}
Another approach to constructing hybrid systems based on SC circuits was proposed by \textcite{Rabl:2006} and extended by \textcite{Rabl:2007}, which consists of atoms and a charge qubit (or a transmon qubit) electrically coupled to a CPW resonator. Afterward, this type of atomic HQC was recognized as a promising approach and most architectures employed similar structures, as shown in Fig.~\ref{HSAS}(a).

In this type of atomic HQC~\cite{Rabl:2006,Rabl:2007,Petrosyan:2009,Tordrup:2008a,Tordrup:2008b,Zhang:2009a,Deng:2010}, the clouds of  atoms (or polar molecules) and charge qubits would be placed at the maximum of the microwave field. Thus, atoms could be positioned parallel to the CPW resonator and longitudinally at the antinode of the electric field, while the charge qubit could be integrated on the ground planes of the CPW resonator also at the antinode of the electric field, see Fig.~\ref{HSAS}(a). Both the atoms and the SC qubit electrically couple to the resonator via microwave fields, which are used as a quantum data bus. The dynamics of such an atomic HQC can be described by Hamiltonians such as those in Eqs. (\ref{SCRE}) and (\ref{TC}).

In contrast with the coupling strength between s SC qubit and the resonator, the electric-coupling strength of an atom is much smaller. In order to achieve strong coupling, atoms (or polar molecules) with large electric dipole moments and suitable frequencies should be employed in this type of atomic HQC. Furthermore, by replacing a single atom with an atomic ensemble, the effective coupling strength between the ensemble and the resonator can become much larger and (depending on the number of atoms) can lie in the strong-coupling regime, which enhances the experimental feasibility. For example, in \textcite{Rabl:2006}, the proposed number of polar molecules in the ensemble was $\sim 10^6$ with the single-molecule coupling strength $g_{\rm atom-cavity}\sim 2\pi\times 10$ kHz. The effective coupling strength is then $g_{\rm eff}\sim 2\pi\times 10$ MHz, which is comparable with that of SC qubits, $g_{\rm sc-cavity}\sim 2\pi\times 10$ MHz. 

Based on this atomic HQC, swap operations between the different components, rotation operations on the atomic ensemble, and entangling operations between different ensembles could be implemented by transferring real photons between two qubits (either atoms or SC qubits) in the resonator and by operations on the SC qubit, which can be dynamically controlled via an external magnetic field threading it, as proposed by \textcite{Rabl:2006}. In this proposal, polar molecule ensembles were used to couple to the resonator by a laser-induced Raman transition because of their relatively large electric-dipole coupling and ability to control the Raman transition. Moreover, quantum states could also be transferred between the atomic system and the SC qubit via virtual photons ~\cite{Zhang:2009a}. Note that flux qubits can also be integrated in atomic HQCs with a CPW resonator. Because a flux qubit would interact with the resonator via magnetic coupling, it should be placed at the antinode of the magnetic field. Thus, if one can experimentally cool and trap the atoms above the CPW resonator, this atomic HQC would be promising to implement novel quantum devices.

\bigskip

\emph{Magnetic-coupling HQCs ---}
Besides electric-dipole coupling, the magnetic dipole moment can also be used to couple atoms to a CPW resonator, while charge qubits electrically couple to the resonator, as shown in Fig.~\ref{HSAS}(b). In this type of atomic HQC, the atoms would be trapped above and parallel to the CPW resonator at an antinode of the magnetic field between the SC wire and ground planes, and the charge qubit would be integrated on the ground planes at an antinode of the electric field. 

For example, \textcite{Verdu:2009} theoretically considered employing a hyperfine transition (at a frequency 6.83 GHz) of $^{87}$Rb atoms, which constitute the atomic ensemble in the HQC. In this proposal, the $^{87}$Rb atoms would be positioned slightly above the gap between the SC wire and the ground planes of the CPW resonator, where the magnetic field is strongest. The dynamics of the interaction between the atoms and the resonator can be described by a Hamiltonian of the form given in Eq.~(\ref{TC}). Consequently, the quantum state could be transferred between the atomic ensemble and the resonator. Furthermore, by integrating a SC qubit into the system, this architecture could also implement various quantum gate operations as in the electric-coupling atomic HQCs.

The magnetic coupling between atoms and the resonator can effectively reduce the effect of charge noise in the system. However, the magnetic-coupling strength is still not sufficiently larger [about 40 kHz in \textcite{Verdu:2009}] than the decay rate $\kappa/2\pi\sim 7$ kHz of the resonator. Thus, it is necessary to increase the resonator's quality factor $Q$ and thereby decrease the photon loss rate. 

\bigskip

\emph{Rydberg atoms in HQCs ---}
Rydberg atoms possessing large electric dipole moments and suitable frequencies can also be used in atomic HQCs with SC qubits. For instance, the proposal of \textcite{Petrosyan:2009} involving a similar structure of the SC circuit as in \textcite{Rabl:2006} [see Fig.~\ref{HSAS}(a)] employed Rydberg states of $^{87}$Rb atoms as the memory of the atomic HQC. 

\begin{figure}[t]
\includegraphics[width=3.35in]{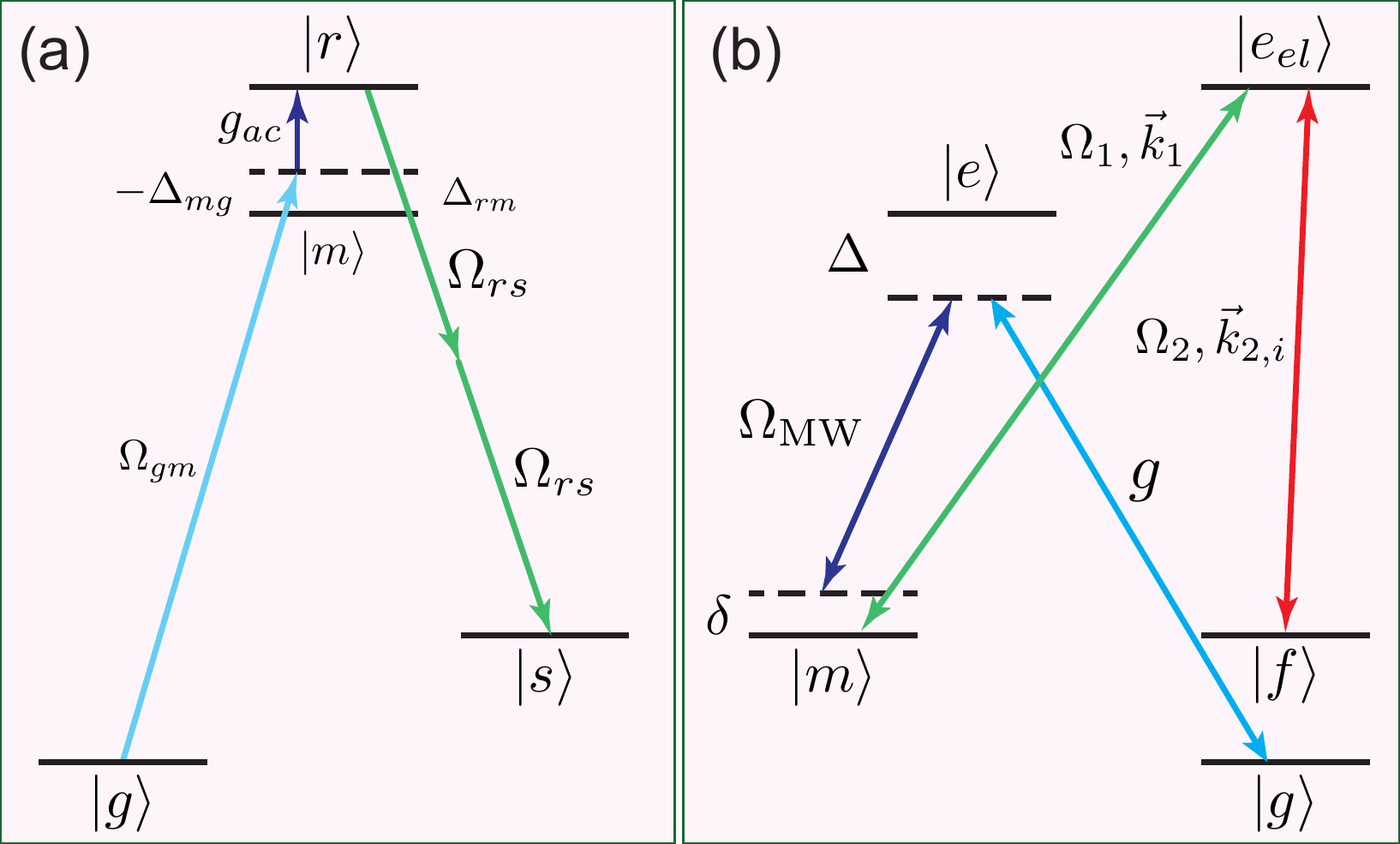}
\caption{(color online). Level structure, driving lasers (with Rabi frequencies $\Omega_{gm}$, $\Omega_{rs}$, $\Omega_{MW}$, $\Omega_1$, and $\Omega_2$), and relevant couplings (with coupling stregths $g_{ac}$ and $g$) to microwave photons in (a) \cite{Petrosyan:2009} model; (b) \cite{Tordrup:2008b} model. (a) and (b) are adapted from \cite{Petrosyan:2009} and \cite{Tordrup:2008b}.}
\label{HAAL}
\end{figure}

In this proposal, the charge qubit, possessing a large coupling strength and good tunability, was used in the circuit. On the atomic system side, the Rydberg states of $^{87}$Rb atoms ($|m\rangle$ and $|r\rangle$) [see Fig.~\ref{HAAL}(a)] were used to interact with the electric field in the resonator. By making a rotating-frame transformation, the coupling between Rydberg atoms and the CPW resonator can be described by the Hamiltonian:
\begin{eqnarray}
H_{ac}=\!&\!\!&\!\hbar\Delta_{mg}m^{\dag}m + \hbar(\Delta_{mg}+\Delta_{rm})r^{\dag}r\nonumber\\
&&\!\!-\hbar(\Omega_{gm}m^{\dag}g+g_{ac}r^{\dag}ma+H.c.),
\end{eqnarray}
where $\Delta_{mg}$ ($\Delta_{rm}$) is the detuning between the frequency of an applied field (the photon in the resonator) and the energy difference of the atomic transition $|g\rangle\leftrightarrow|m\rangle$ ($|m\rangle\leftrightarrow|r\rangle$) ($\Delta_{rm}\simeq-\Delta_{mg}=\Delta$ in this proposal), see Fig.~\ref{HAAL}(a). The operators $m$ ($m^{\dag}$), $r$ ($r^{\dag}$), and $g$ ($g^{\dag}$) annihilate (create) an atom in state $|m\rangle$, $|r\rangle$, and $|g\rangle$, respectively, and $a$ ($a^{\dag}$) is the photon annihilation (creation) operator. The Rabi frequency of this applied laser field and the coupling strength of the resonator's photon with the atom are $\Omega_{gm}$ and $g_{ac}$, respectively. When a photon is generated in the resonator, the laser field and the photon would excite the atoms from the ground state $|g\rangle$ to the Rydberg state $|r\rangle$ via a nonresonant intermediate Rydberg state $|m\rangle$, which is a two-photon transition. In the regime when $\Delta_{mg}\gg g_{ac},\Omega_{gm}$, the population of state $|m\rangle$ can be neglected, and the transition between states $|g\rangle$ and $|r\rangle$ is described by
\begin{equation}
V_{gr}=\hbar g_{\rm eff}(r^{\dag}ga+a^{\dag}g^{\dag}r)
\end{equation}
with $g_{\rm eff}=\Omega_{gm}g_{ac}/\Delta$. Then, by introducing another laser field with Rabi frequency $\Omega_{rs}$, the atoms can be transferred from the Rydberg state $|r\rangle$ to the storage state $|s\rangle$ via a two-photon transition. 

Thus, the process of transferring a quantum state from the SC qubit to this type of Rydberg atoms could be implemented as follows: First, the charge qubit is brought to resonance with the CPW resonator for a duration of time $\tau_{\rm SC}=\pi/(2g_{\rm SC})$ (SC qubit $\rightarrow$ resonator); next, the laser field with $\Omega_{gi}$ is turned on for a time duration $\tau_{gr}=\pi/(2\sqrt{N}g_{\rm eff})$ (resonator $\rightarrow$ Rydberg state $|r\rangle$ of atoms); finally, the laser field with $\Omega_{rs}$ is switched on for a time duration $\tau_{rs}=\pi/(2\Omega_{rs})$ (Rydberg state $|r\rangle$ of atoms $\rightarrow$ storage state $|s\rangle$). A similar sequence can be used for the opposite process.

The operation times [$\tau_{gr},\tau_{rs}\simeq1~\mu$s in \textcite{Petrosyan:2009}] are much smaller than the relative decay times in this atomic HQC, and the quantum information could therefore be transferred between the SC and atomic qubits. However, auxiliary optical fields can also be absorbed by the SC electrodes of the CPW resonator before driving the atomic transition. A thin metallic mirror can be added to cover the SC electrodes as designed in \textcite{Petrosyan:2009}, but reductions in the resonator's quality factor $Q$ and the fidelity of the system may occur. This is a problem that should also be solved.

\bigskip

\emph{Many-qubit storage and computation ---}
In addition to single-channel information processing, atomic HQCs can, in principle, be used for many-qubit storage and computation. \textcite{Tordrup:2008a} and \textcite{Tordrup:2008b} proposed two theoretical methods to achieve this purpose by using an atomic HQC containing an ensemble of polar molecules and a charge qubit. 

\textcite{Tordrup:2008a} proposed to store qubit states in different rotational excited states of the molecular ensemble by using Raman transitions. \textcite{Tordrup:2008b} proposed to store qubit states in different collective-excitation modes of plane-wave form. In both of these proposals, quantum gate operations are implemented using an integrated charge qubit.

The proposed storage and retrieval method used in \textcite{Tordrup:2008b} is similar to some of the ideas that are being pursued in more recent proposals and experiments, and we explain it in more detail. In addition to two hyperfine rotational ground states $|g\rangle$ and $|f\rangle$ for qubit storage, an auxiliary rotational ground state $|m\rangle$ and two rotationally and electronically excited states $|e\rangle$ and $|e_{el}\rangle$ are utilized, as shown in Fig.~\ref{HAAL}(b). The storage process can be described as follows: First, the information is stored in the auxiliary collective state $|m,0\rangle=(1/\sqrt{N})\sum_i|g_1\cdots m_i\cdots g_N\rangle$ via a Raman transition; then through a stimulated Raman adiabatic passage process with classical optical fields, the information is transferred to the collective state $|f,\vec{q}_i\rangle=(1/\sqrt{N})\sum_ie^{i\vec{q}_i\cdot \vec{x}_i}|g_1\cdots f_i\cdots g_N\rangle$, with momentum $\vec{q}_i=\vec{k}_1-\vec{k}_{2,i}$. The dynamics of this process is described by the Hamiltonian: 
\begin{eqnarray}
H=\sum_i\!\!\!&&\!\!\!\left(\Omega_1e^{i\vec{k}_1\cdot\vec{x}_i}|e_{el,i}\rangle\langle m_i|\right.\nonumber\\
\!\!\!&&\left.+\;\Omega_2e^{i\vec{k}_{2,i}\cdot\vec{x}_i}|e_{el,i}\rangle\langle f_i|+H.c.\right),
\label{TORDRUP}
\end{eqnarray}
where $\Omega_1\exp\{i\vec{k}_1\cdot\vec{x}_i\}$ and $\Omega_2\exp\{i\vec{k}_{2,i}\cdot\vec{x}_i\}$ describe two classical optical fields. The interaction in the above equation could be used to encode a pattern $\exp\{i\vec{q}_i\cdot \vec{x}_i\}$ in the collective state $|f,\vec{q}_i\rangle$ from the auxiliary collective state $|m,0\rangle$. The retrieval process could be implemented by reversing the storage procedure. By employing appropriate encoding methods of storing and retrieving quantum information as introduced in \textcite{Tordrup:2008b}, in principle, many-qubit storage could be implemented in such a circuit.

To perform a single-qubit operation in the SC circuit, the relevant state should be transferred to the SC qubit through the CPW resonator. Two-qubit gates could also be realized by transferring one state to the SC qubit and another to the resonator first, and then operating on the state via the interaction between the SC qubit and the CPW resonator.

\subsubsection{Spin hybrid quantum circuits}\label{sssec:shqc}

\begin{figure}[t]
\includegraphics[width=3.04in]{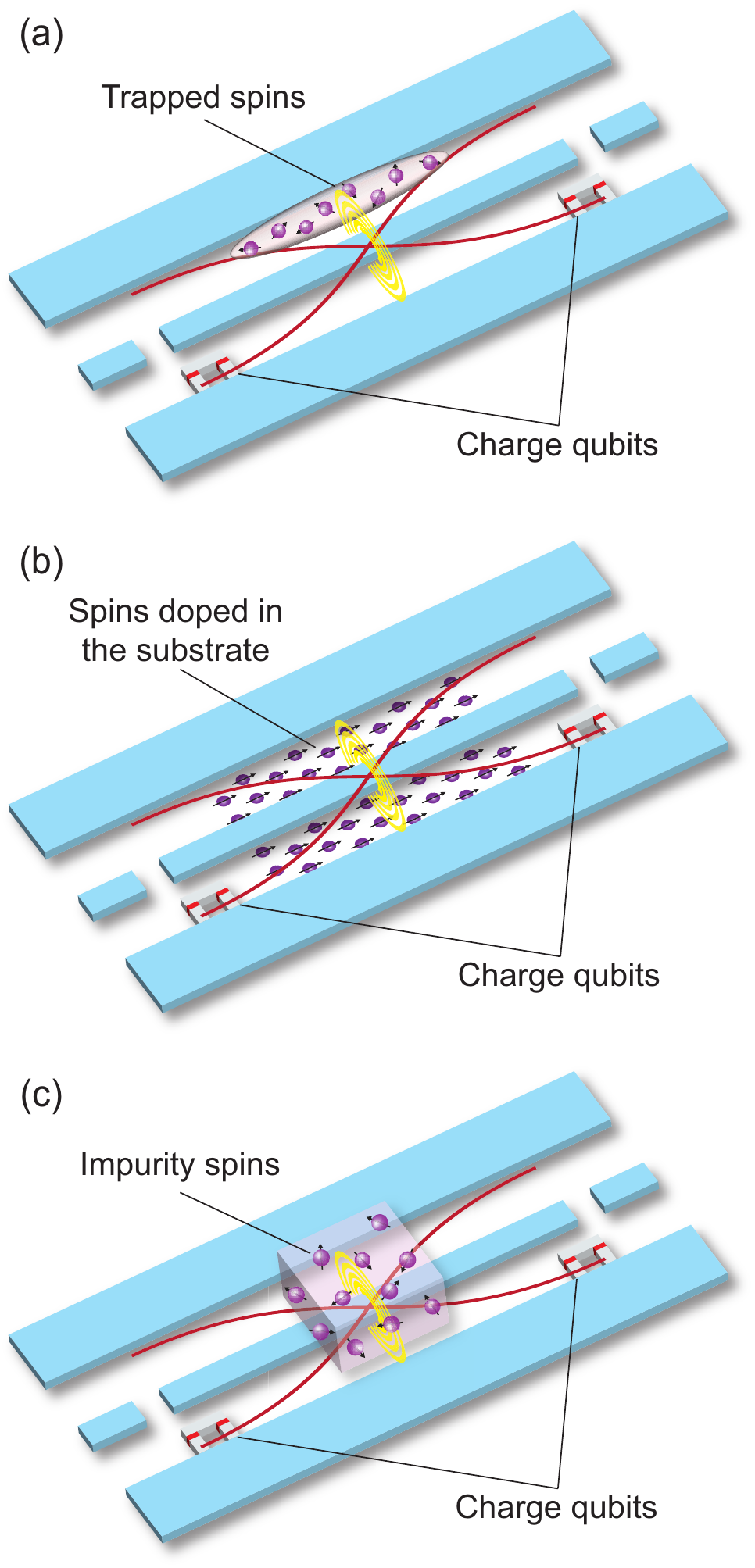}
\caption{(color online). Schematic diagrams of three types of hybrid systems combining ensembles of spins and superconducting resonators. (a) The ensemble of spin-carrying atoms is trapped using electromagnetic fields. (b) The ensemble of spins is doped on the surface of the substrate of the superconducting resonator. (c) The ensemble of spins is fixed in a sample of diamond or ruby that is placed on top of the superconducting wire of the resonator. As in Fig.~\ref{HSAS}, superconducting qubits are also coupled to the coplanar waveguide resonator, so the resonator can be used as a data bus to indirectly couple the spins and SC qubits.}
\label{HSSS}
\end{figure}

Facing the challenge of improving cooling and trapping techniques required for atomic HQC, the focus recently shifted to ensembles of spins, as shown in Fig.~\ref{HSSS}, which can also serve as quantum memories with long coherence times and are much easier to integrate in a solid-state device than atoms, because spins can be doped in the device and do not require any complicated trapping techniques or large electromagnetic fields to bring them in resonance with the CPW resonator~\cite{Imamoglu:2009,Wesenberg:2009,Yang:2011a,Yang:2011b,Ping:2012}. With SC qubits, this spin HQC can be used to implement quantum gate operations as mentioned in atomic HQCs and simultaneously be more conveniently integrated into a small chip in experiment. 

Unlike most atomic ensembles, spin ensembles couple to the CPW resonator via their magnetic dipole moment instead of the electric dipole moment. They should therefore be placed at the antinode of the magnetic field in the resonator in order to achieve maximum coupling, as shown in Fig.~\ref{HSSS}. As an essential advantage of this design, this hybrid system is insensitive to charge noise, and thus long coherence times could be achieved. The charge qubit (or the transmon) is still placed at the antinode of the electric field as in atomic HQCs.

\bigskip

\emph{Hybrid two-level system ---}
\textcite{Imamoglu:2009} proposed a method to construct an effective two-level system from a spin ensemble coupled resonantly to a transmon qubit through a CPW resonator. In this proposal, the spins of the electrons in the substrate or cold ground-state atoms trapped above the resonator are used as the ensemble of spins to be placed at an antinode of the resonator's magnetic field, as shown in Fig.~\ref{HSSS}. The resonator is strongly coupled to the transmon qubit and thereby behaves as a nonlinear cavity. For strong coupling between the spin ensemble and the resonator, a hybrid two-level system is obtained by choosing appropriate parameters of the coupling strength and the detuning of the spin ensemble.

\bigskip

\emph{Holographic quantum register ---}
\textcite{Wesenberg:2009} proposed using the electron spin of nitrogen atoms in fullerene cages (N@C$_{60}$), with a suitable frequency in the microwave range, as the memory medium in the CPW resonator to construct the hybrid system with a transmon qubit. A large number of N@C$_{60}$ are doped into or deposited on the surface of the substrate and a transmon qubit is integrated at an antinode of the electric field, as shown in Fig.~\ref{HSSS}(b). 

On the side of the memory, by using the collective excitation of the spin ensemble, this system can also have a large coupling strength with the resonator, whose dynamics is described by a Hamiltonian of the form given in Eq.~(\ref{TC}). Furthermore, one could employ a magnetic field gradient $(z\hat{z}-y\hat{y})\Delta B/L$ to the substrate for a duration $\tau$, where $\hat{z}$ is along the CPW resonator and $\hat{y}$ is perpendicular to the plane of the resonator. These conditions provide a spatially varying Zeeman energy shift $e^{ikz}$, which results in a spatially varying phase shift operation. This operation moves the stored quantum information between the different collective-excitation modes in the spin ensemble. These modes are defined by the excited states
\begin{equation}
|e(k)\rangle\equiv\frac{1}{\sqrt{N}}\sum_i\frac{g_i}{\bar{g}}e^{ikz_i}|g_1\cdots e_i\cdots g_N\rangle,
\end{equation}
where $g_i$ is the coupling strength of the spin at position $\mathbf{r}_i$, and $\bar{g}\equiv\sqrt{\sum_i|g(\mathbf{r}_i)|^2/N}$ denotes an averaged coupling strength; $k$ is the wave number and serves as an index for the different modes. A number of collective-excitation modes of the same spin ensemble can therefore be used as many channels to store quantum information by appropriately choosing the magnetic field gradient pulses. The storage and retrieval of new data does not disturb previously stored data in different spin modes, because only the $k=0$ spin mode interacts with the field in the resonator. 

In addition, this proposal also could implement single- and two-qubit gate operations with the transmon qubit~\cite{Tordrup:2008a,Tordrup:2008b}. By transferring quantum information from the target qubit formed in a spin ensemble to the SC qubit via the CPW resonator, the single-qubit operation could be implemented on the SC qubit. Moreover, one can implement two-qubit gates by swapping quantum information from two modes in the spin ensemble to the SC qubit and the resonator and then implementing a two-qubit gate operation between the SC qubit and the resonator.

\emph{Other spin hybrid quantum circuits ---}
A recent proposal~\cite{Ping:2012} demonstrated that a hybrid circuit, whose main component is a spin ensemble coupled to a CPW resonator, can in principle be used for performing measurement-based quantum computing. A SC qubit and a second resonator are used in the proposal, but only for the purpose of implementing the measurements in the protocol.

In two other proposals, HQCs that integrate ensembles of NV centers, CPW resonators and SC qubits were investigated. In the proposal of \textcite{Yang:2011b}, a phase qubit is employed as a quantum processor, while spin ensembles act as quantum memories. Various quantum operations, such as the preparation of multiqubit $W$ states in the collection of the spin ensembles can be performed using the SC qubit. \textcite{Yang:2011a} proposed using a phase qubit, which is capacitively coupled to two CPW resonators, to entangle two spin ensembles placed in these two resonator. In this proposal, the resonators act as data buses and the phase qubit plays the role of a tunable coupler.

\section{Experimental realization of hybrid systems with spins and superconducting circuits}\label{sec:erhqc}

Besides theoretical proposals on HQCs, significant progress has been made on the experimental realization of HQCs involving spin ensembles and SC qubits. A strong coupling between spin ensembles and SC resonators was achieved first. Afterwards, a direct-coupling hybrid circuit (NV centers and a flux qubit) as well as an indirect-coupling hybrid circuit (NV centers and a transmon qubit in a SC resonator) were also experimentally realized.

\subsection{Direct-coupling hybrid circuit with nitrogen-vacancy centers and a flux qubit}\label{ssec:dcnv}

\begin{figure}[t]
\includegraphics[width=3.3in]{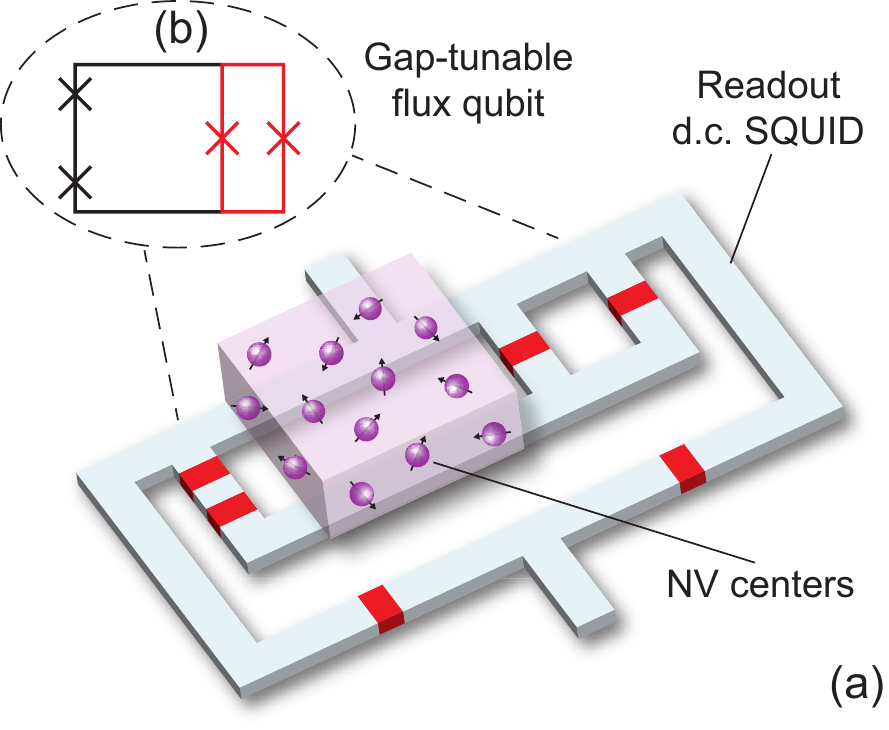}
\caption{(color online). (a) Sketch of the experimental setup in \textcite{Zhu:2011}. A diamond crystal (an ensemble of NV centers) was glued on top of a superconducting circuit, where a gap-tunable flux qubit [shown in (b)] (two small loops in the center) and a readout dc SQUID (the largest loop) shared a common edge. (b) Circuit diagram of the gap-tunable flux qubit used in the central part of the superconducting circuit shown in (a).}
\label{FLNVE}
\end{figure}

The theoretical proposal studied by \textcite{Marcos:2010} (see Sec.~\ref{ssec:dcc}) was recently realized experimentally~\cite{Zhu:2011}. In this experiment, a sample of diamond containing $\sim 3\times10^7$ NV centers was glued on top of the SC circuit [Fig.~\ref{FLNVE}(a)], which consisted of a flux qubit [Fig.~\ref{FLNVE}(b)] and a readout dc SQUID [the largest loop in Fig.~\ref{FLNVE}(a)] inductively coupled to the qubit. For better tunability, they used a low-inductance dc SQUID loop to replace the smallest of the three Josephson junctions in the flux qubit. Thus, by controlling the magnetic fluxes threading the main loop of the flux qubit [the left, big loop in Fig.~\ref{FLNVE}(b)] and the small loop of the d.c. SQUID [the right, small loop in Fig.~\ref{FLNVE}(b)], the energy splitting of the flux qubit can be adjusted. Note that this experiment differed from the proposal of \textcite{Marcos:2010} in that the states $|m_s=\pm1\rangle$ of NV centers were not split by the weak external field, but rather hybridized by the strain-induced field. As a result, a superposition of these two states is involved in the coupling to the flux qubit.

From the spectroscopic measurements of the flux qubit coupled to the spin ensemble, a vacuum Rabi splitting was observed and the coupling strength between the two systems reached 70 MHz. Thus, this direct-coupling HQC could be used to transfer quantum information between the two components. By tuning the flux qubit into resonance with the spin ensemble, single-energy-quantum exchange between the two systems was observed with the decay time $\sim 20$ ns. 

\begin{figure*}[t]
\includegraphics[width=5.3in]{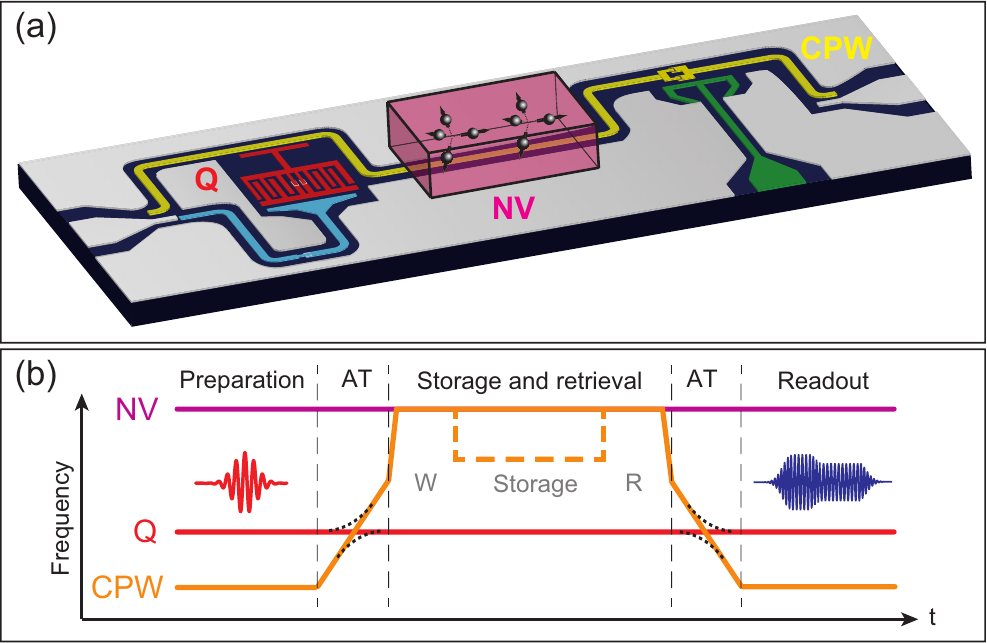}
\caption{(color online). (a) Sketch of the experimental device in \textcite{Kubo:2011}. (b) Schematic diagram of the preparation, storage, retrieval, and readout processes. In panel (b), the letters AT denote adiabatic transfers of the quantum state between the SC qubit and the coplanar waveguide resonator. The letters W (write) and R (read) refer, respectively, to the transfer of the quantum state to and from the spin ensemble (i.e., the memory). Note that the write-store-read sequence was not implemented in the experiment. In both panels, the letters Q, NV, CPW refer, respectively, to the transmon qubit, the NV centers in diamond, and the coplanar waveguide resonator.}
\label{KUBO}
\end{figure*}

The most likely sources of decoherence are the large electron-spin bath of $P$1 centers (a nitrogen atom substituting a carbon atom), which are generated when the sample is prepared. Because the two types of electron spins naturally couple to each other in the experiment~\cite{Zhu:2011}, applying an external magnetic field to change the energy gap of NV centers or using a better sample could effectively improve the decay time of the NV center ensemble. Another source of decoherence is the strong hyperfine interaction between the NV center spins and $^{13}$C nuclear spins, which is approximately 100 MHz. This source of decay might be reduced by polarizing the nuclear spins. Therefore, in principle, the decoherence can be reduced and better direct-coupling HQCs can be experimentally realized.

\subsection{Spins coupled to superconducting resonators (without qubits)}\label{ssec:scsr}

In order to achieve an indirectly-coupled HQC, one has to have a strong coupling between the SC resonator and both the SC qubit and the spin ensemble. Several experiments demonstrated the possibility to strongly, even ultrastrongly, couple SC qubits with resonators~\cite{Wallraff:2004,Devoret:2007,Sillanpaa:2007,Hofheinz:2008,Hofheinz:2009,Niemczyk:2010}. However, similar experiments on spin ensembles have not been implemented until recently and these have become the first challenge to experimentally realizing HQCs. Note that semiconductor quantum dots have also been experimentally integrated into superconducting resonators~\cite{Frey:2012}; however, their coherence times are much shorter than impurity spins.

Recently, four independent groups experimentally achieved strong coupling between an ensemble of impurity spins and a CPW resonator~\cite{Kubo:2010,Schuster:2010,Bushev:2011,Amsuss:2011,Sandner:2012,Kubo:2012}. They placed a solid-state ensemble of spins on the SC wire at an antinode of the standing wave of the current on the SC wire, where the resonator's magnetic field has a maximum, as shown in Fig.~\ref{HSSS}(c). NV centers in diamond were used by \textcite{Kubo:2010,Kubo:2012}, \textcite{Amsuss:2011}, and \textcite{Sandner:2012}, Cr$^{3+}$ spins in ruby and $P$1 centers in diamond were utilized by \textcite{Schuster:2010}, and Er$^{3+}$ ions in a Y$_2$SiO$_5$ crystal were employed by \textcite{Bushev:2011}. The frequencies of these impurity spins are all compatible with those of SC resonators. In order to achieve strong coupling between the spins and the resonator, an ensemble of $10^{12}$ electron spins was used to enhance the coupling strength by about 6 orders of magnitude. Thus, the coupling strength between the ensemble and the resonator reached 10--65 MHz. The vacuum Rabi splitting, photon exchange frequency, and even the storage and retrieval of a microwave field were observed in these experiments, providing evidence that the exchange of microwave photons between the ensemble of spins and the CPW resonator can indeed take place.

In a matter-cavity hybrid system, the loss from the matter and the cavity is commonly described by a measure called the cooperativity, which is defined as $C=g^2/\kappa\gamma$, where $g$ is the qubit-resonator coupling strength, while $\kappa$ and $\gamma$ are the decay rates of the cavity and the spins (or atoms, etc.). In \textcite{Kubo:2010}, \textcite{Schuster:2010}, and \textcite{Bushev:2011}, the cooperativities were 7, 27, and 11.5, respectively. This means that the coupling strength between the spins and the CPW resonator is in the strong-coupling regime, and photons can be coherently transferred into the spin ensemble. Recently, \textcite{Huebl:2012} reported a spin-cavity hybrid system with a higher cooperativity ($\sim 1350$). In this experiment, an ensemble of $\sim4.5\times10^{16}$ spins in gallium-doped yttrium iron garnet was strongly coupled to the cavity mode of the CPW resonator. The coupling strength was $\sim 450$ MHz, which is about $13\%$ of the frequency of the resonator. 

In order to achieve longer coherence times in spin ensembles, \textcite{Wu:2010} and \textcite{Amsuss:2011} attempted the possibility of transferring energy from the electron spins to the nuclear spins by using the hyperfine interaction between them~\cite{Childress:2006,Dutt:2007,Jiang:2008,Sar:2012}. In principle, this would improve the coherence times from several hundred microseconds in the electron spins to seconds in the nuclear spins. In addition, \textcite{Wu:2010} demonstrated that a spin ensemble can simultaneously store a number of different microwave modes by using magnetic field gradients.

\begin{table*}
\label{Table3}
{\small
\begin{center}
\colorbox{mybo}{
\begin{tabular}{@{}p{3.7cm}p{3.3cm}p{1.9cm}p{2.2cm}p{2.25cm}p{2.1cm}p{1.5cm}@{}}
\multicolumn{7}{l}{\rule[-3mm]{0mm}{8mm}{\bfseries TABLE III ~Coupling strengths between resonators and other systems in different proposals and experiments.}}\\
\hline\hline
\rule[-3mm]{0mm}{8mm}Reference & Atoms or spins & Coupling mechanism & Number of $~~~$ atoms or spins &  Coupling to $~~~~~~~$ atoms or spins & Coupling to $~~~$ SC qubit& Cooper- $~$ativity\tablenotemark\\
\hline
\rule[-3mm]{0mm}{8mm}{\it Theoretical Proposals} & & & & & &\\
\cite{Rabl:2006} & Molecules & Electric & $10^4$--$10^6$ & 1--10 MHz & $\lesssim50$ MHz & $>10^3$\\
\cite{Tordrup:2008b} & Atoms or Molecules & Electric & $10^5$--$10^6$ & 1--10 MHz & 200 MHz & $ >10^3$\\
\cite{Verdu:2009} & Atoms ($^{87}$Rb) & Magnetic & $\sim10^6$ & $40$ kHz & 50 MHz & 1.7\\
\cite{Petrosyan:2009} & Rydberg atoms ($^{87}$Rb) & Electric & $\sim10^6$ & 3.85 MHz & 50 MHz & $1.5\times10^4$\\
\cite{Imamoglu:2009} & Spins & Magnetic & $\sim10^8$ & 10 MHz & $>100$ MHz & $3.9\times10^3$\\
\cite{Wesenberg:2009} & Spins (N@C$_{60}$) & Magnetic & $\sim10^{11}$ & 6 MHz & $\sim100$ MHz & $>4\times10^2$\\
\hline
\rule[-3mm]{0mm}{8mm}{\it Experiments} & & & & &\\
\cite{Kubo:2010} & Spins (NV centers) & Magnetic & $\sim10^{12}$ & 11.6 MHz & --- & 27 \\
\cite{Schuster:2010} & Spins (Cr$^{3+}$ or N) & Magnetic & $\sim10^{12}$ & 65 MHz & --- & 11.5 \\
\cite{Bushev:2011} & Spins (Er$^{3+}$) & Magnetic & $\sim10^{12}$ & 20 MHz & --- & 7 \\
\cite{Amsuss:2011} & Spins (NV centers) & Magnetic & $\sim10^{12}$ & 10 MHz & --- & 10--20\\
\cite{Kubo:2011} & Spins (NV centers) & Magnetic & $\sim10^{11}$ & 3--4 MHz & 7 MHz & 15--20 \\
\hline\hline
\end{tabular}}
\end{center}
}
\tablenotetext{The cooperativities listed here are for the coupling between a CPW resonator and either atoms or spins; for the coupling between a CPW resonator and a SC qubit, the cooperativity is usually larger than $10^3$.}
\end{table*}

\subsection{Indirect-coupling hybrid circuits with nitrogen-vacancy centers and a transmon qubit}\label{shqce}

Almost simultaneously with the experiment of \textcite{Zhu:2011}, another experiment achieved coupling between an ensemble of NV centers and a transmon qubit using a SC resonator as an intermediary~\cite{Kubo:2011}. The experiment is described in Fig.~\ref{KUBO}. In this setup, a diamond crystal consisting of $\sim 10^{11}$ NV centers was placed at the center of the resonator where the magnetic field has a maximum, and a transmon type qubit was placed on one side of the resonator, where the electric field has a maximum. The coupling strength between the spin ensemble (the transmon) and the resonator was $\sim 3$ MHz (7.2 MHz).

In order to couple these two systems with different frequencies, a tunable CPW resonator was used. In this resonator, a SQUID was embedded in order to make the frequency of the resonator tunable (through its dependence on the applied magnetic flux threading the SQUID loop). Thus, storage, readout and transfer of quantum information can be achieved by tuning the frequency of the resonator into resonance with the SC qubit and the spin ensemble. In the experiment, these processes were experimentally realized as follows [Fig.~\ref{KUBO}(b)]: First, after preparing a quantum state in the SC qubit, the frequency of the resonator was adiabatically swept across the frequency of the qubit, transferring the qubit state to the corresponding photonic state. This process is more immune to flux noise in the SQUID than putting the two systems in resonance for a carefully set duration~\cite{Wei:2008}. The frequency of the resonator was then brought into resonance with the spin ensemble for some duration. The quantum state then oscillates between the resonator and the spin ensemble. The frequency of the resonator was then tuned away from that of the spin ensemble, and adiabatically swept across the frequency of the qubit, transferring the photonics state of the resonator to the SC qubit. Finally, the state of the SC qubit was measured. In principle, a storage and retrieval process could be implemented by adjusting the interaction time between the resonator and the spin ensemble in such a way to exactly swap the quantum state between the two systems, as shown by the dashed line in Fig.~\ref{KUBO}. For technical reasons, however, this experiment was not performed~\cite{Kubo:2012a}. 

However, in the presence of interference effects caused by the hyperfine structure of NV centers and  the inhomogeneous broadening at resonance, the fidelity, which describes the correspondence of the readout signal with the original signal, was very low (about $10\%$) and the coherence times were not very long (about several hundred nanoseconds). Higher-purity diamond could greatly improve the performance of NV centers (long coherence). However, one has to keep in mind that a high concentration of spins is desirable for the purpose of achieving strong coupling. Thus, one of the challenges at the moment is to find the best compromise between high concentration (strong coupling) and diluted impurities (long coherence). There are also efforts aimed at finding methods for enhancing the coherence of these systems, such as dynamical decoupling~\cite{Lange:2010,Huang:2011,Naydenov:2011}.

\bigskip

\begin{figure*}[t]
\includegraphics[width=7in]{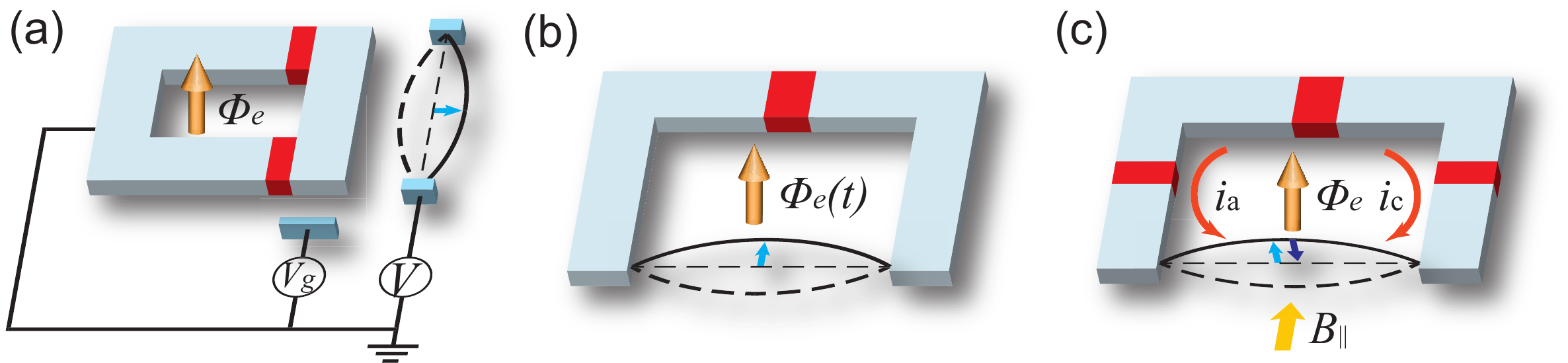}
\caption{(color online). Schematic diagrams of hybrid quantum circuits consisting of nanomechanical resonators (represented by the solid and dashed lines) and (a) a superconducting charge qubit, (b) an rf-SQUID, and (c) a three-junction flux qubit.}
\label{NANOS}
\end{figure*}

Table III lists some parameters, including coupling strengths between different systems and CPW resonators and related cooperativities. These parameters are essential for achieving an effective exchange of quantum information between the quantum bus (the resonator) and both the memory (the atoms or spins) and the processor (the SC qubits). All these results of combining atomic ensembles (including spins) and CPW resonators to build hybrid circuits point toward more progress in the future.

\section{Hybrid quantum circuits with nanomechanical resonators}\label{sec:nrsq}

There are three main approaches that have been considered for coupling a SC qubit and a NAMR: via a neighboring capacitance [see Fig.~\ref{NANOS}(a)], by changing an applied magnetic flux [see Fig.~\ref{NANOS}(b)], or via a Lorentz force induced by a loop current [see Fig.~\ref{NANOS}(c)].

Recently, many groups have devoted considerable effort to this type of HQC. In such a HQC, the NAMR can couple to a SC circuit and serve as a cavity. The entire configuration thereby provides a solid-state analog of cavity-QED systems, and it can reach the strong-coupling regime~\cite{Schwab:2005}. Meanwhile, the NAMRs are generally much smaller in size than CPW resonators, and thus they can be more easily integrated in high-density quantum devices. Various designs and applications of quantum (or nonclassical) behavior in this type of HQC have been studied, such as the generation of quantum entanglement~\cite{Armour:2002,Cleland:2004}, quantum measurement~\cite{Lahaye:2009}, high precision displacement detection~\cite{Etaki:2008}, and cooling~\cite{Wei:2006,Grajcar:2008a,Xue:2007b,Teufel:2011,Taylor:2011}.

\subsection{Coupling mechanisms}

\subsubsection{Capacitive coupling}

In this case, the NAMR is capacitively coupled to a charge qubit via a capacitance that depends on the displacement $x$ of the harmonic oscillator, as shown in Fig.~\ref{NANOS}(a). The amplitude of the oscillation of the NAMR is much smaller than the equilibrium distance $d$ between the SC qubit and the NAMR. Thus, the capacitance between them approximately becomes $C(x) \simeq C_0(1-x/d)$, where $C_0$ is the capacitance of the NAMR in equilibrium. This $x$-dependent capacitance leads to a situation where the harmonic-oscillation mode of the NAMR effectively couples to the quantum state of the charge qubit.

\subsubsection{Magnetic flux coupling}

In this case, the NAMR is embedded in the loop of an rf SQUID or a flux qubit and oscillates in the plane of the loop, as shown in Fig.~\ref{NANOS}(b). The displacement of the NAMR leads to a small change in the loop area and therefore the magnetic flux $\Delta\Phi= \alpha Blx$, where $\alpha$ is a geometric factor and $l$ is the length of the cantilever. The total magnetic flux threading the rf-SQUID or the flux qubit becomes an $x$-dependent magnetic flux $\Phi_e[x]=BA+\alpha Blx=\Phi_{\rm eq}+\Delta\Phi[x]$, where $\Phi_{\rm eq}$ is the magnetic flux when the cantilever is at its equilibrium position and $\Delta\Phi[x]= \alpha Blx$ is the flux variation due to the vibration of the cantilever. Because the SQUID and the flux qubit are sensitive to the applied magnetic flux, such an $x$-dependent magnetic flux can induce an effective strong coupling between the quantized harmonic-oscillation mode of the NAMR and the quantum states of the SQUID or the flux qubit.

\begin{figure*}[t]
\includegraphics[width=7in]{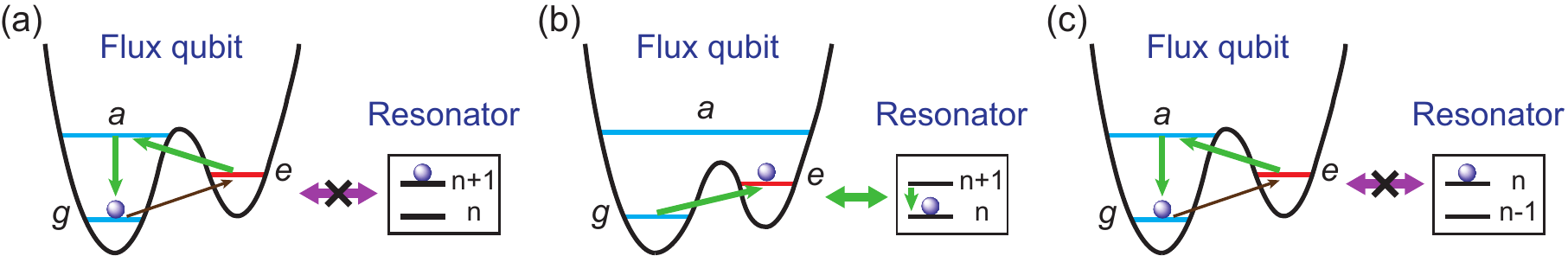}
\caption{(color online). Cooling process in a hybrid quantum system: (a) Cooling brings the flux qubit (on the left) to the ground state while the qubit is off resonance with the resonator (inside the square box on the right). (b) When the flux qubit on the left is switched to resonantly interact with the resonator, the latter is cooled by exciting the qubit to the state $|e\rangle$. (c) Tuning the qubit on the left off resonance from the resonator, the qubit is cooled down again.}
\label{COOL}
\end{figure*}

\subsubsection{Electromotive coupling}
In this case, the NAMR is also embedded in the loop of a flux qubit, but oscillates along the direction perpendicular to the plane of the loop, as shown in Fig.~\ref{NANOS}(c). When an external magnetic field $B_{\|}$ is applied parallel to the plane of the loop, the persistent current can induce a Lorentz force with opposite directions for the clockwise and anticlockwise current states. Meanwhile, the oscillations of the NAMR are modulated by these Lorentz forces. Thus, the quantized harmonic-oscillation mode of the cantilever is coupled to the quantum state of the flux qubit.

\subsubsection{Coupling dynamics}

In general, the dynamics of the three circuits shown above can all be described by the Hamiltonian:
\begin{eqnarray}
H =\!\!&&\!\!\frac{\varepsilon}{2}\,\sigma_z^{\prime}+t\,\sigma_x^{\prime} + \hbar\,\omega_{\rm NAMR}\,b^{\dag}b\nonumber\\
\!\!&&\!\!+\hbar\,g_{\rm SC-NAMR}\left(b^{\dag}\,+\,b\right)\sigma_z^{\prime},
\label{ONANO}
\end{eqnarray}
where $\varepsilon$ and $t$ are the energy difference and the tunnelling amplitude between two states of the SC qubit, respectively, $\vec{\sigma}^{\prime}$ denotes the Pauli operator of the SC qubit, $\omega_{\rm NAMR}$ is the fundamental vibrational mode frequency of the NAMR, $b^{\dag}$ and $b$ are the resonator creation and annihilation operators, respectively, and $g_{\rm SC-NAMR}$ describes the coupling strength between the SC circuit and the NAMR. When the SC qubit works at the degeneracy point with $\varepsilon=0$, by employing the rotating-wave approximation, the Hamiltonian~(\ref{ONANO}) can be reduced to
\begin{eqnarray}
H_{\rm eff}= \!\!&&\!\!t\,\sigma_z+\hbar\,\omega_{\rm NAMR}\,b^{\dag}b\nonumber\\
\!\!&&\!\!+ \hbar \,g^{\ }_{\rm SC-NAMR}\,(b^{\dag}\sigma^-+\sigma^+b),
\label{NANO}
\end{eqnarray}
where $\vec{\sigma}$ denotes the Pauli operators of the SC qubit in the eigenstate basis.

\subsection{Applications}

The physics of the nanomechanical hybrid circuits introduced above can be described by a Jaynes-Cummings model, as in Eq.~(\ref{NANO}), which is a close analog to cavity QED. Hence, many applications of quantum devices built in other cavity-QED systems can also be implemented in nanomechanical HQCs~\cite{Armour:2002,Cleland:2004,Xue:2007a,Lahaye:2009,Liu:2010,Chen:2011,Shevchenko:2012}. Futhermore, the NAMR can be small in size, which facilitates the HQCs' scalability. In addition, a CPW resonator can also be coupled to a nanomechanical hybrid circuit to form a new quantum device~\cite{Sun:2006,Didier:2011,Pirkkalainen:2012}. For instance, \textcite{Sun:2006} designed a quantum transducer. In this proposal, a charge qubit is designed to capacitively couple with the NAMR and magnetically couple to the CPW resonator simultaneously. Quantum information can therefore be coherently exchanged between the CPW resonator and the NAMR by controlling the charge qubit and implement the quantum analog of the transducer used in classical telephones.

Besides serving as a cavity, another promising application of nanomechanical HQCs is to observe the quantum and classical behaviors of NAMR oscillations. If a NAMR with sufficiently high oscillation frequency ($\sim$ GHz) is cooled to very low temperatures ($\sim$ mK) (i.e., the quantum oscillation energy becomes larger than the thermal energy $k_BT$), then the NAMR can approach the quantum limit and exhibit various quantum phenomena~\cite{O'Connell:2010,Teufel:2011}. Many approaches have been proposed to achieve the cooling of the oscillator with optomechanical systems~\cite{Marquardt:2009}, where an oscillating micro-mirror or cantilever is proposed as a harmonic oscillator. Alternative approaches for cooling NAMRs can be found in nanomechanical hybrid circuits~\cite{Zhang:2005,Xue:2007b,You:2008,Zhang:2009b,Xia:2009,O'Connell:2010,Teufel:2011}.

For instance, in a capacitive-coupling nanomechanical hybrid circuit, as shown in Fig.~\ref{NANOS}(a), the charge qubit can serve as a coolant that absorbs energy from the NAMR~\cite{Zhang:2005}. In such a system, the cooling procedure consists of two parts: a relaxation process (off resonant) and a cooling process (resonant). By varying the external magnetic flux threading the qubit loop, one can control the transition between these two processes, which switch the interaction on and off between the qubit and the NAMR, and the relaxation rate of the qubit. In the relaxation process, the charge qubit is switched far off resonance with the resonator and taken to a fast-decay bias point. The interaction between the two parts can therefore be neglected and the state of the qubit can relax to the ground state. In the cooling process, by adjusting the external flux to switch on the interaction between the charge qubit and the NAMR, the energy can be transferred from the NAMR to the qubit in an appropriate cycle interval. Repeating these two processes, energy is continuously extracted from the NAMR, and thus the NAMR is cooled, possibly close to its ground state where its quantum features become apparent.

Furthermore, besides cooling the NAMR, the whole system (including the solid-state circuit part) can also be cooled down simultaneously~\cite{You:2008}, which is useful to enhance the quantum coherence properties in the quantum device. \textcite{You:2008} proposed such a HQC involving a four-junction flux qubit, where the small Josephson junction is replaced by a tunable SQUID. The cooling process can be described as follows: first, the energy levels of the flux qubit are adjusted as shown in Fig.~\ref{COOL}(a), where the qubit is off resonance with the resonator and the transition rates satisfy the relations $\Gamma_{ag}>\Gamma_{ea}\gg\Gamma_{ge}$. The noise from the outside environment can excite the qubit to its first excited state. After optically pumping the qubit from the first to the second excited state, the qubit can later on quickly decay to the ground state, due to the large transition rate $\Gamma_{ag}$. This process ensures that the qubit remains close to the ground state~\cite{Grajcar:2008b,Valenzuela:2006}. Then, by adjusting the magnetic field threading the qubit loop, the qubit is switched to resonantly interact with the resonator for a period of time [see Fig.~\ref{COOL}(b)]. This process extracts energy from the resonator. Afterward, the qubit is switched back off resonance and cooled down to the ground state again as above [see Fig.~\ref{COOL}(c)]. Repeating these processes, both the qubit and the resonator can be simultaneously cooled.

Recent rapid progress made in nanoscience has stimulated the design of different models to use HQCs to achieve the cooling of a quantum device and a coupled nanomechanical resonator. Such cooled HQCs provide a promising platform for exploring various quantum phenomena and for implementing the quantum-to-classical transition in a macroscopic system.

In addition, NAMRs can also act as a bridge linking atoms or spins with SC quits. Recently, two theoretical designs provided two different approaches to implement such a HQC, where the NAMR simultaneously connects the SC circuit with a Rydberg atom~\cite{Gao:2011} or a NV center~\cite{Chen:2010}.

\section{Other hybrid quantum circuits}\label{sec:oqsc}

\subsection{Hybrid quantum circuits with microscopic defects}\label{ss:tlsjj} 
\begin{figure}[t]
\includegraphics[width=3.4in]{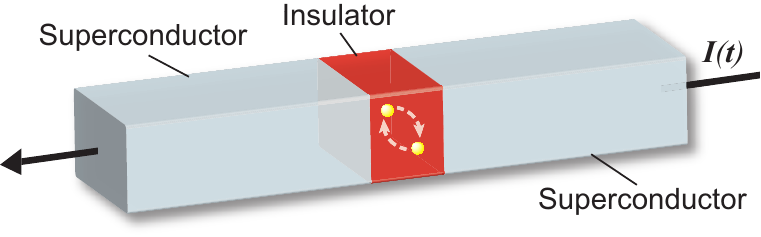}
\caption{(color online). Schematic diagram of two-level systems inside the oxide tunnel barrier of a Josephson junction.}
\label{TLS}
\end{figure}

Besides the above atomic or spin HQCs, microscopic TLS defects that naturally occur in Josephson junctions can also interact with SC qubits and constitute a new type of hybrid circuits. Spurious TLSs inside the amorphous oxide tunnel barrier of junctions can be atoms or electrons having two possible positions inside potential wells with tunneling between them, as shown in Fig.~\ref{TLS}. Generally, TLSs are regarded as a nuisance because they cause decoherence~\cite{Martin:2005,Ku:2005}. However, it was proposed~\cite{Zagoskin:2006}, and experimentally demonstrated~\cite{Neeley:2008,Sun:2010} that these TLSs, which sometimes have long coherence times, can also be used as quantum memories. This approach provides a new method to utilize TLS defects in solid-state devices.

\textcite{Zagoskin:2006} theoretically analyzed the quantum TLSs in current-biased Josephson junctions (CBJJ) and these were studied as qubits and also as quantum memories. The dynamics of the TLS-CBJJ system can be described by the following effective Hamiltonian:
\begin{equation}
H=\frac{\hbar \omega}{2}\sigma_z+\sum_j\left(\frac{\Delta_j}{2}\tilde{\sigma}_z^j+\lambda_j\sigma_x\tilde{\sigma}_x^j\right),
\end{equation}
where the Pauli matrices $\vec{\sigma}$ ($\vec{\tilde{\sigma}}$) operate on the two lowest states of the CBJJ (the TLS states), $\omega$ is the interlevel spacing of the CBJJ, $\Delta_j$ denotes the interlevel spacing of the $j$th TLS, and the coupling strength between the CBJJ and the TLS is described by the coefficient $\lambda_j$. In this Hamiltonian, the last term $\lambda_j\sigma_x\tilde{\sigma}_x^j$ leads to energy exchange between the CBJJ and the TLS when they are resonant with each other. This means that the storage and retrieval of quantum information in TLSs can be implemented when the CBJJ is switched to resonance with the TLSs for an appropriate operation time. Other relevant theoretical and experimental proposals of operating TLSs~\cite{Tian:2007,Neeley:2008,Lisenfeld:2010,Sun:2010} have been reported, and these suggest that this type of HQC could have better properties for coherent quantum storage by using TLSs rather than SC qubits.

\subsection{Hybrid quantum circuits with topological qubits}\label{ssec:tq} 
Topological quantum computation is currently being explored by various groups [see \textcite{Nayak:2008} for a review]. A topological quantum system has topologically-ordered states of matter that are insensitive to local perturbations. Thus, this quantum system can be protected from decoherence when it serves as a qubit. However, coherently coupling topological qubits with other qubits in a common hybrid system would prove challenging because this coupling is very weak~\cite{Vishveshwara:2011}.

Recently, \textcite{Bonderson:2011} and \textcite{Jiang:2011} theoretically studied two different interfaces between topological qubits and other conventional qubits. \textcite{Jiang:2011} proposed to couple a topological qubit with a SC flux qubit, and \textcite{Bonderson:2011} designed a scheme to allow a topological qubit to interact with a semiconductor double-dot qubit via a SC flux qubit. In both proposals, a pair of exotic quasiparticles (Majorana fermions) was used as the topological qubit, and was localized on the segments formed by a topological insulator~\cite{Jiang:2011}, or on a wire whose conduction electrons have strong spin-orbit coupling~\cite{Bonderson:2011}. The Majorana fermion is an exotic particle which is its own antiparticle. This pair of Majorana fermions is in a nonlocal quantum state and possesses non-Abelian quantum statistics, which are crucial for topological quantum computation.

\textcite{Jiang:2011} theoretically considered the proximity effect~\cite{Tinkham:1996} to link the macroscopic wave function of the superconductor with the Majorana fermions at the interface between the superconductor and topological insulator via the tunneling of Cooper pairs between these two materials~\cite{Fu:2008}. Consequently, the superconducting phase, which depends on the state of the flux qubit, can coherently control the Majorana fermions in the topological qubit and be designed as a key ingredient of controlled-phase gate to achieve the exchange of quantum information between the SC qubit and the topological qubit. \textcite{Bonderson:2011} theoretically considered the Aharonov-Casher effect to effectively couple the topological qubit to a double quantum dot qubit~\cite{Hanson:2008a}. The Aharonov-Casher effect is dual to the familiar Aharonov-Bohm effect, and it states that a neutral particle possessing a magnetic moment, such as a vortex in a superconductor, can obtain a quantum phase shift while moving around a line charge. When the Majorana fermion pair and quantum dots are placed on the flux qubit, the Aharonov-Casher effect makes the state of the flux qubit sensitive to the electron parity of the Majorana fermion pair and quantum dots. Thus, the parity measurements with the help of the flux qubit could be used to entangle qubits and coherently transfer quantum information.

Moreover, a method to coherently transfer quantum information and perform other quantum operations on topological qubits and spins in quantum dots was also theoretically proposed recently~\cite{Leijnse:2011,Leijnse:2012}.

\subsection{Hybrid quantum circuits for converting optical photons to microwave photons}

Most proposals and experiments introduced above involve microwave photons in the GHz frequency range. However, there are many quantum systems working in the visible or infrared frequency range. Recently, \textcite{DiVincenzo:2011} proposed a hybrid superconductor-optical quantum repeater which integrates an optical system of a visible or infrared frequency with a SC circuit in the GHz range. In this quantum repeater, the SC and optical subsystems couple to each other via a microwave transmission medium, such as a CPW. The process of exchanging a quantum state between them could be implemented as follows: First in the optical subsystem, a photon is received and transferred via an optical channel. Then it is downconverted to a microwave photon in the microwave transmission medium. Finally the quantum signal is transferred from the microwave photon and stored or operated on the SC qubit. This process can also be applied in the opposite direction.

In such a HQC, the key issue in transferring the quantum signal between optical and SC subsystems is the down-conversion process of the optical photon to the microwave photon, and the inverse process (i.e., up-conversion)~\cite{Strekalov:2009}. \textcite{DiVincenzo:2011} proposed using a nanoscale tunnel junction, which has a nonlinear current-to-voltage characteristic, to link the optical and microwave signals. This nonlinear tunnel junction can convert the optical power to a current source as
\begin{equation}
I(\omega=0)=\eta E(\omega_{\rm opt})E^*(\omega_{\rm opt}),
\end{equation}
where $I(\omega=0)$ is the output dc current,  $\eta$ is the efficiency coefficient, and $E(\omega_{\rm opt})$ the electric field at the tunnel junction, which has a frequency $\omega_{\rm opt}$. In the downconversion process, there are two optical fields introduced at the junction. One is the signal photon from the optical subsystem with frequency $\omega_{\rm opt} +\Delta\omega$, and the other is the strong laser beam with frequency $\omega_{\rm opt}$. The output current produced by the nonlinear junction is $I(\Delta\omega) =\eta E(\omega_{\rm opt})E^*(\omega_{\rm opt} +\Delta\omega)$, where $\Delta\omega$ can be in the GHz range, matching the working frequency of a SC subsystem. As a result, quantum information carried in the optical photon can be converted to a current signal with GHz frequency and then transferred as a microwave photon in the microwave transmission medium. In the opposite process,  the output current produced by the nonlinear junction becomes $I(\omega_{\rm opt}+\Delta\omega)=\eta E(\omega_{\rm opt})E^*(\Delta\omega)$, and the microwave photon can then be up-converted to the optical photon. For other related work, see \textcite{Ilchenko:2003}, \textcite{Matsko:2007}, \textcite{Tsang:2010}, \textcite{Tsang:2011}, \textcite{Wang:2012}, and \textcite{Tian:2012}.

\section{Discussion and conclusion}

Hybrid quantum circuits promise to open up new possibilities for quantum technologies where the hybrid circuit combines the best features of its constituents, such as long coherence times, fast operations and scalability.

We presented an overview of the current status of HQCs, especially devices integrating atomic or spin systems with SC qubits or resonators, where considerable progress has been made recently in both theory and experiment.

Even though the basic ideas for coupling the different systems together (e.g., relying on the natural coupling to electric or magnetic fields) are quite simple, there are a variety of different approaches for putting a hybrid circuit together. For example, one can decide to use direct coupling between atomic and solid-state qubits or use a SC resonator to mediate the coupling by balancing the requirements of setup simplicity and sufficient coupling strength. Moreover, there are many elements that can be used as the atomic or spin part of the hybrid device, and careful consideration of the properties of the different candidates must be considered before deciding which one is best-suited for achieving the desired purpose of the device.

Experiments on HQCs are just starting to demonstrate the coherent coupling between different physical systems. However, with rapid progress in fabrication techniques, one can expect better results in the coming few years. In particular, the coupling of NV centers in diamond to SC qubits and resonators promises to be an active area of research for years to come, and one can envision such a HQCs to be a central component of quantum technologies ranging from precision-measurement devices to quantum computers.

Hybrid devices involving NAMRs are also starting to enter the quantum regime. With a constantly increasing level of control, a number of experiments have demonstrated quantum effects in such HQCs in the past few years, and these circuits are now becoming increasingly feasible as components in technologies that involve mechanical sensing devices.

The field of hybrid circuits is still evolving, with new ideas emerging steadily. It is quite likely that in a few years the designs will be significantly more sophisticated than the prototypes that have been studied in recent years.

\section*{Acknowledgements}

We thank Sahin Ozdemir for a careful reading of the manuscript and useful suggestions. We also thank Iulia Georgescu, Xuedong Hu, Yuimaru Kubo, Sergey Shevchenko, and Anders S\o rensen for useful discussions and comments. SA and FN were supported in part by the ARO, JSPS-RFBR Grant 12-02-92100, Grant-in-Aid for Scientific Research (S), MEXT Kakenhi on Quantum Cybernetics, and the JSPS through the FIRST program. ZLX and JQY were supported in part by the National Basic Research Program of China Grant No.2009CB929302, NSFC Grant No.91121015, and MOE Grant No.B06011. ZLX was also partly supported by the RIKEN IPA program.

\end{document}